\documentclass[manuscript]{acmart}

\AtBeginDocument{%
  }

\setcopyright{acmlicensed}
\copyrightyear{2025}
\acmYear{2025}
\acmDOI{XXXXXXX.XXXXXXX}

\usepackage{adjustbox}
\usepackage{pdfpages}

\newcommand{\change}[1]{\textcolor{black}{#1}}

\newcommand{\rev}[1]{\textcolor{black}{#1}}

\begin{document}

\title[Theory of Troubleshooting: Developer's Experience of Overcoming Confusion]{Theory of Troubleshooting: \\The Developer's Cognitive Experience of Overcoming Confusion}

\author{Arty Starr}
\email{artystarr@uvic.ca}
\orcid{0009-0007-4025-1516}
\affiliation{%
  \institution{University of Victoria}
  \city{Victoria}
  \state{BC}
  \country{Canada}
}

\author{Margaret-Anne Storey}
\email{mstorey@uvic.ca}
\orcid{0000-0003-2278-2536}
\affiliation{%
  \institution{University of Victoria}
  \city{Victoria}
  \state{BC}
  \country{Canada}
}

\begin{abstract}
\rev{This paper introduces a Theory of Troubleshooting that is rooted in cognitive science. This theory helps software developers explain the challenges they face and the project risks that emerge as troubleshooting becomes difficult. We define \textit{troubleshooting} as the cognitive problem-solving process of identifying, understanding, and constructing a mental model of the cause of an unexpected system behavior, and consider the cognitive process of troubleshooting to be an integral part of the activity of debugging. Troubleshooting is a particularly intense and draining aspect of software work, placing sustained demands on attention, working memory, and mental modeling.  By surfacing and naming the \textit{confusion experience} inherent in troubleshooting in terms of neurological and attentional dynamics, our theory explains how prolonged troubleshooting can deplete cognitive resources and lead to cognitive fatigue.  In the study presented in this paper, we interview 27 professional developers about their troubleshooting experiences, and follow a Constructivist Grounded Theory approach to construct a theory grounded in empirical data.  Our theory contributes to research on Developer Experience by providing a cognitive foundation for understanding troubleshooting difficulty, fatigue, and sustainability risk—and offers practical implications for both research and industry.}

\end{abstract}




\maketitle

\section{Introduction}

Improving Developer Experience has become an important goal for many companies as a strategy to improve productivity and employee retention.  However, the lack of visibility and difficulty in quantifying the problems \rev{that developers face} create barriers in communication with management, and often prevents the investment of time and resources that would make Developer Experience better \cite{greiler_actionable_2023}.  Many developers resort to coping strategies such as reducing engagement, no longer speaking up, or leaving their job as a last resort, if they are unable to improve the working conditions \cite{greiler_actionable_2023}.  For managers that lack experience of what software coding is like \cite{Kalliamvakou_great_manager2019}, it is not clear why the developers are getting so frustrated.

In a keynote on the Neuroscience Behind Developer Productivity Engineering, Hans Dockter—\change{founder of Gradle—highlighted a persistent communication breakdown: developers often have a strong intuitive sense of when their tools, systems, or processes are creating drag, but struggle to explain this ``gut feel'' to managers who do not share the lived experience of writing code \cite{gradle_dpe_2022}.  This disconnect makes it difficult to justify improvements in Developer Experience, even when those improvements could enhance productivity and unlock innovation capability.  Dockter proposed that \textit{cognitive fatigue} is a key concept from cognitive science that could help bridge this gap—offering a way to make the developer’s intuition more understandable to others.  He argued that the software industry urgently needs better conceptual foundations grounded in cognitive science—frameworks that can explain the developer’s experience in a way that does not require having lived it firsthand.  Based on this insight, he introduced Developer Productivity Engineering (DPE): an engineering discipline aimed at reducing cognitive fatigue and improving productivity, including a core focus on reducing troubleshooting time \cite{gradle_dpe_2022}}.

\rev{The most closely related area of software engineering research to troubleshooting is debugging. While the terms \textit{troubleshooting} and \textit{debugging} are often used interchangeably, they are not exact synonyms. They describe overlapping processes with different orientations: \textit{troubleshooting} focuses on the cognitive problem-solving process of understanding unexpected system behavior and is more related to the developer's cognitive experience, while \textit{debugging} refers to the overall activity of removing bugs from the program \cite{layman_debugging_2013}, which involves the cognitive process of troubleshooting.
We define \textit{troubleshooting} as the cognitive problem-solving process of identifying, understanding, and constructing a mental model of the cause of an unexpected system behavior, and consider troubleshooting (cognitive process) to be an integral part of the activity of debugging.}

While cognitive fatigue has been widely studied in domains outside software development \cite{ackerman_cognitive_2011}, research within this context remains limited.  One exception is a survey by Sarkar and Parnin (n=311), which found that a majority of developers experience severe and frequent issues with cognitive fatigue \cite{sarkar_characterizing_2017}. \rev{Their} findings highlight the need to better understand the cognitive demands placed on developers and how they contribute to fatigue.  \rev{In software development, such demands are not evenly distributed across activities. Troubleshooting, in particular, places sustained demands on attention, working memory, and mental modeling, as developers work to reconcile unexpected system behavior with their existing understanding. DPE’s emphasis on reducing troubleshooting time as a strategy to reduce cognitive fatigue \cite{gradle_dpe_2022} points to an intuitive recognition of troubleshooting as a particularly intense and draining aspect of software work—one that may offer valuable insight into the invisible struggles developers face.}

\rev{In the study presented in this paper, we interviewed 27 professional developers about their troubleshooting experiences, and followed a Constructivist Grounded Theory (CGT) approach \cite{charmaz_constructing_2014} to construct a Theory of Troubleshooting grounded in empirical data.} As a central research question, we asked: ``What is the developer thinking, feeling, and striving for during the experience of troubleshooting?'' \rev{Our theory offers a conceptual foundation rooted in cognitive science that helps developers explain the challenges they face as troubleshooting becomes more difficult. By surfacing and naming the \textit{confusion experience} in terms of neurological and attentional dynamics, the theory explains how prolonged troubleshooting can deplete cognitive resources and lead to fatigue.  This has project-level implications: when software complexity rises and troubleshooting becomes difficult, it introduces a sustainability risk. We propose that designing software systems for easier troubleshooting is a practical strategy for managing these risks. } 

Our \rev{theory} is designed with \rev{utility} in mind: rather than offering abstract or overly technical descriptions, we emphasize natural language that reflects how developers themselves describe their experiences. In doing so, \rev{we aim to provide resonant language} that helps developers articulate their internal experiences, and enables clearer conversations with managers, tool designers, and researchers seeking to improve Developer Experience. While this is a theory-building study and does not include a system implementation or intervention, our aim is to support practical use of our theory in both research and industry contexts.

\rev{The paper begins with a review of related work in debugging, cognitive fatigue, and Developer Experience.} Next, we describe our methodology, outlining the CGT approach and data analysis process. In the findings section, we present our Theory of Troubleshooting, introducing a set of experience-centered models that capture how developers think, feel, and strive during troubleshooting. The discussion explores the implications of \rev{our theory} for understanding cognitive fatigue, improving Developer Experience, and informing risk management and tool design. We conclude by summarizing our contributions and outlining directions for future work.

\section{Background}

\rev{Although much has been written about debugging and Developer Experience, we still lack a grounded understanding of the developer's internal experience of troubleshooting—particularly when something feels wrong but cannot yet be explained. Prior work typically focuses on observable behaviors \cite{perscheid_studying_2017}, tool interactions \cite{beller_dichotomy_2018}, or broad emotional states \cite{wrobel_emotions_2013}, with some studies investigating neurological processes \cite{krueger_neurological_2020}. In contrast, our study explores how developers describe their experiences by asking open-ended questions reflecting on specific troubleshooting experiences and difficulties.  Using a CGT approach, we construct a theory with an experience-oriented paradigm that incorporates thinking, feeling, and striving factors.  Rather than tracking debugging behavior in real-time with an observational study, we focus on the developer’s cognitive experience as remembered and interpreted—how developers understand and describe what was happening for them during these moments.}

As background context, we begin by reviewing prior research on the process of debugging, including empirical observations and \rev{existing} models.  Next, we examine research on cognitive fatigue, which motivated our focus on troubleshooting and plays a central role in our investigation.  Finally, because our goal is to \rev{understand} the developer's experience, we review existing definitions, frameworks, and key findings from Developer Experience research to ground our theoretical perspective.

\subsection{Software Debugging}

One definition of debugging \rev{given by Layman et al.} is: “To detect, locate, and correct faults in a computer program” \cite{layman_debugging_2013}.  A bug is an error (or fault) in a program that causes the program’s behavior to be inconsistent with the developer’s or user’s expectations \cite{shaochun_xu_cognitive_2004}.  Debugging is a complex activity that involves a demanding cognitive process \cite{shaochun_xu_cognitive_2004}\rev{, that we identify in this paper as troubleshooting.}

Although we distinguish between troubleshooting and debugging, \rev{these terms are} of course closely related, and the literature of debugging also sheds light on the experience of troubleshooting.  We begin our review by examining prior software engineering research on debugging, in particular we focus on 1) findings from observational studies of real-world debugging, 2) lab studies \rev{that reveal insights from fMRI's, eye-tracking, and controlled experiments of debugging}, and 3) \rev{models of debugging} similar to our own to help position our study within the context of existing theoretical research.

\subsubsection{Observing Real-World Debugging}

\change{Some of the most compelling insights into debugging come from empirical studies where developers encounter real bugs in the wild.  These studies offer a grounded view of debugging not as a clean, formal process, but as a messy, intuitive, and often tacit practice.}  \change{In a study by Perscheid et al., developers were asked to think aloud while troubleshooting bugs during their regular workday \cite{perscheid_studying_2017}.}  When confronted with unexpected behavior, one developer first checked which code was recently modified, then set breakpoints at key locations of the program flow to interrupt the program and check the program state. Another developer examined the log files to get an impression of the system parts involved. Several developers found it difficult to explain their debugging approach, but described the overall process as a simplified version of a scientific process, or an intuitive method.

Beller et al. collected real-world usage data from 458 developers by instrumenting IDEs with a plugin and \rev{conducted} surveys \cite{beller_dichotomy_2018}. They found widespread use of print statements to debug and were surprised to see that developers did not use the debugger at all in 91\% of IDE sessions. Beller's findings are consistent with Perscheid's finding that there is lagging adoption of advanced debugging tools and features, and developers have little to no education on debugging \cite{beller_dichotomy_2018}.

Alaboudi and LaToza investigated debugging in the wild by studying live-streamed programming videos from YouTube and Twitch, where 11 professional developers worked on open source projects \cite{alaboudi_what_2023}.  \rev{They} identified two types of debugging contexts, 1) \rev{fixing} bugs reported in an issue tracker, and 2) \rev{bugs} inserted or triggered during the programming task, which they called ``debugging episodes''.  \change{In studying 89 debugging episodes, durations ranged from five seconds to two hours, and they observed that developers continuously edit and run their code, a pattern they called ``edit-run cycles'' \cite{alaboudi_edit_2021}. Developers would edit and run their code an average of seven times before fixing a defect.}

In another study involving real bugs, \change{Böhme et al.} conducted an experiment with twelve professional software engineers who together spent one month on localizing, explaining, and fixing 27 real bugs in several widely used programs \cite{bohme_how_2017}.  For each bug, the developers got a small and succinct bug report, the source code containing the bug and executable, and a test case that failed because of the bug.  The developers were asked to 1) point out the buggy program statements (\textit{fault localization}), 2) explain how the error comes about (\textit{bug diagnosis}), and 3) develop a patch (\textit{bug fixing}).  

\change{Böhme et al.'s} findings show that developers provide essentially the same explanation for an error, providing empirical evidence that bug diagnosis is not a subjective matter \cite{bohme_how_2017}.  The developers also perceived that four of the 27 bugs were ``very difficult'' to diagnose because certain flags, functions, or data structures were left undocumented.  \change{Their rationale—that the difficulty stemmed from undocumented code—suggests that the developers were struggling to infer the intentions of prior developers who wrote the original code, which impeded program comprehension.}  \change{The process of trying to infer the intentions of previous developers is often overlooked in models of debugging \cite{gilmore_models_1991}}.

\subsubsection{Observing Developers in the Lab} \label{lab}

\rev{Lab studies enable fine-grained analysis of debugging behavior under controlled conditions, allowing researchers to examine developer strategies, visual attention, and cognitive effort. Two recent studies by Vaithilingam et al. (n=24) \cite{vaithilingam_expectation_2022} and Barke et al. (n=20) \cite{barke_grounded_2023} investigate the lived experiences of developers using GitHub Copilot, revealing important shifts in the developer's experience related to debugging. Both studies show a common pattern: developers found it easier to get started with the AI tools, but then \textbf{faced increased difficulties when troubleshooting} the generated code.  If troubleshooting and cognitive fatigue are connected, then this shift of increasing troubleshooting difficulty could have broader implications for developer experience (increasing fatigue).}

Lab studies \rev{using} fMRI or fNIRS to monitor developers' brain activity during debugging offer a window into the neurological processes \rev{of} software development. When combined with eye-tracking, these methods enable researchers to \rev{identify what developers are looking at, and correlate visual attention with brain activity.}  Krueger et al. (n=30) \cite{krueger_neurological_2020} found that, compared to writing prose, writing code activates more of the brain's right hemisphere, engaging regions associated with attention control, working memory, planning, and spatial cognition.  Hu et al. (n=28) \cite{hu_towards_2024} used fNIRS and eye-tracking in a study focused specifically on debugging and confirmed similar activation patterns during code editing, and further identified distinct neural activation signatures at \rev{several} stages of the debugging process.

\change{Duraes et al.\ (n=15) \cite{duraes_wap_2016} combined fMRI and eye-tracking to investigate developers’ brain activity while identifying bugs in source code, focusing on the particular moment when developers first suspect they have found a bug.  At this point, they observed increased activation in the insula, a brain region involved in decision-making under uncertainty, attentional control, and executive function.  This neural activation suggests that the suspicion of a bug may coincide with an onset of uncertainty—initiating a shift in cognitive mode toward evaluating, deciding, and reorienting attention in response to a puzzling or ambiguous event. This interpretation aligns closely with patterns that emerged in our analysis, which we developed independently of this prior work.}

\subsubsection{Models of Debugging}

\change{Several researchers have proposed models to describe the process of debugging.} Katz and Anderson offered one of the earliest, presenting debugging as a four-step sequence: comprehend the system, test the system, locate the error, and fix the error \cite{katz_debugging_1987}. While influential, this model assumes that developers build a complete understanding of the system before beginning fault localization.  Gilmore challenged this assumption by distinguishing between the “comprehension process” and the “state of comprehension,” arguing that understanding unfolds dynamically throughout the debugging process \cite{gilmore_models_1991}. 

\begin{figure*}[h]
\centerline{\includegraphics[scale=0.5]{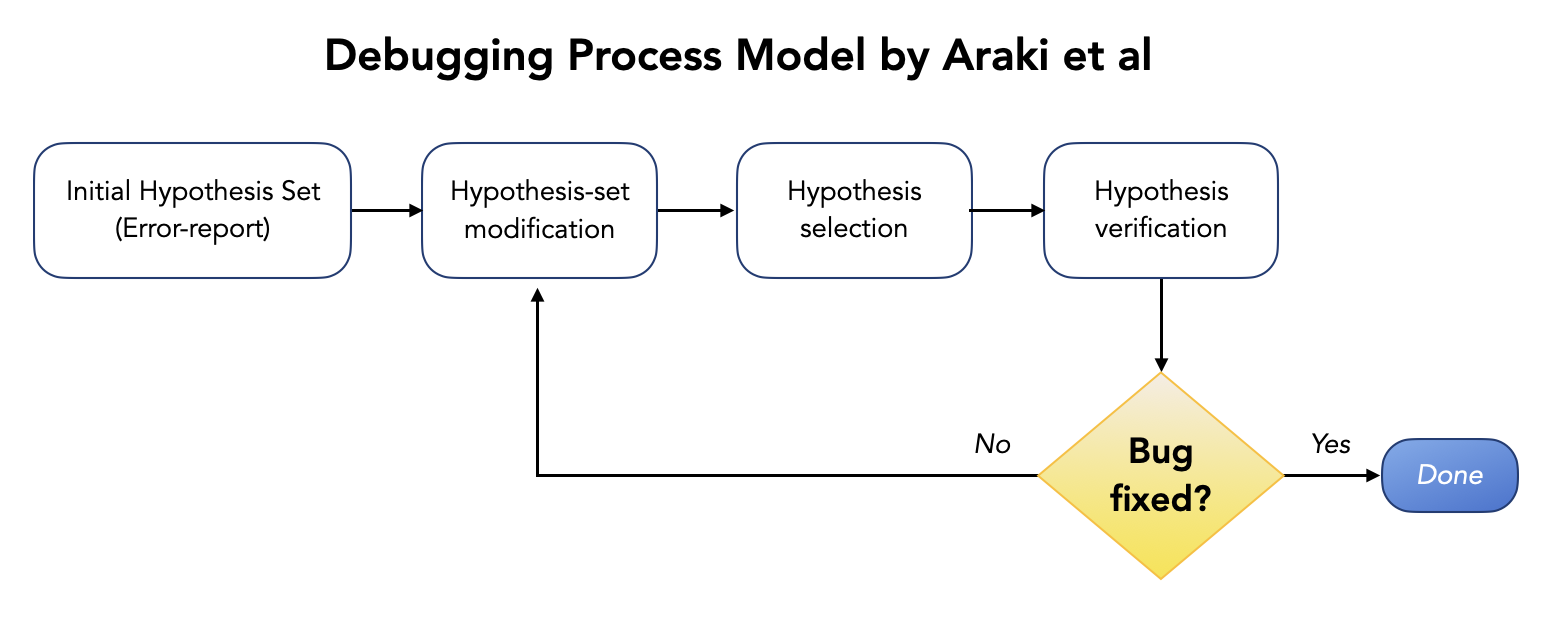}}
\caption{Redrawn debugging process model by Araki et al. showing a hypothesis-driven process \cite{araki_general_1991}}
\label{fig:hypothesis-loop}
\end{figure*}

\change{Several other models characterize debugging as an iterative hypothesis-driven process.  Araki et al. characterizes debugging as an iterative process of developing hypotheses and verifying or refuting them \cite{araki_general_1991} \rev{as} shown in Figure \ref{fig:hypothesis-loop}.}  As debugging proceeds, developers modify and refine the hypothesis set, and the hypothesize-validate loop is repeated until the bug is found.  A number of studies characterize the process of debugging using a similar hypothesis-validate loop model, for example, \change{Li et al.} describes, \textit{``Typically, developers would study the failure at hand, make hypotheses on what the cause(s) of the failure may be, and examine a subset of the execution that can confirm or disprove their hypothesis. They would then leverage the additional understanding of the failure acquired in this process to make new hypotheses or refine the existing ones, examine additional subsets of the execution, and so on. This process would continue until either the developers give up or they find the fault.''} \cite{li_enlightened_2018}  \rev{Both of these characterizations of debugging center the cognitive process aspects that we refer to as troubleshooting.}

Vessey studied the debugging behavior of novice and expert developers, finding that the process does not always revolve around hypothesis testing \cite{vessey_expertise_1985}. Experts often began without a hypothesis, and still took \rev{meaningful} debugging actions. \rev{In several cases}, even after \rev{articulating} a hypothesis, the developers did not actively evaluate it, instead allowing their understanding of the problem to unfold \rev{as they gained familiarity.} Vessey observed that expert developers were formulating a mental model of the \rev{expected correct system behavior, and perceiving the error as a deviation from correct behavior, without forming a hypothesis about the underlying cause \cite{vessey_expertise_nodate}. This contrasts with Araki’s hypothesis-driven loop, highlighting a different driver of action—where developers respond to a felt misalignment between actual and expected behavior and navigate by intuitive sense, rather than explicit causal reasoning.}

\rev{While these models offer valuable insight into the structure of debugging activity, they do not capture the developer's experience during debugging, or explain why the activity can be particularly frustrating, mentally taxing, or fatiguing. By centering the developer's cognitive experience and relating developers’ experiential accounts to mechanisms described in cognitive science, we offer a new lens for understanding debugging and the factors that contribute to cognitive fatigue.}

\subsection{Cognitive Fatigue}

\change{Most research on cognitive fatigue has been conducted outside the domain of software development \cite{ackerman_cognitive_2011}, leaving a gap in how its findings relate to developers’ lived experience. In this section, we summarize core concepts from the cognitive fatigue literature that are most relevant to software development, particularly during complex troubleshooting.  While some parallels are evident, more nuanced connections between developers' experiences and the causes of fatigue will be addressed in the findings (Section~\ref{findings}), where we introduce theoretical constructs grounded in our data.}

\subsubsection{Performance and Compensating Effort}

Historically, cognitive fatigue research has focused on performance during extended engagement with demanding tasks, measuring changes in reaction time, error patterns, and response quality over time \cite{ackerman_cognitive_2011}. However, this \rev{performance-centered} approach has important limitations, as performance outcomes are shaped by a complex interplay of factors. For example, individuals experiencing fatigue often report an urge to stop or reduce effort, but can override it by exerting compensatory effort to stay focused \cite{ackerman_cognitive_2011}. \rev{This compensating effort refers to the extra mental or physical exertion required to sustain performance, even when internal resources are depleted.}

Although this compensating effort may preserve short-term performance, it carries physiological costs. \rev{Research shows that} exerting such compensatory effort can increase the release of stress hormones (e.g., cortisol, noradrenaline), elevate blood pressure, and amplify the experience of fatigue \cite{robert_j_hockey_compensatory_1997}. In high-stakes environments, such as software development under tight deadlines, these responses may temporarily mask performance deficits while contributing to chronic stress and burnout.

\subsubsection{Controlled vs. Automatic Processing Tasks} \label{controlled_vs_automatic processing}

\change{Cognitive fatigue does not affect all types of mental activity equally.  Tasks that require cognitive control because of novel, inconsistent, or ambiguous information are more susceptible to fatigue-related decrements because they demand sustained attention and deliberative effort \cite{schneider_controlled_1977}.  Such tasks are prevalent in software development. In contrast, automatic processing tasks place lower demands on attention and are less sensitive to fatigue effects \cite{schneider_controlled_1977}.} 

When fatigued, individuals performing controlled tasks tend to exhibit an overall pattern of action disorganization \cite{myers_conceptions_1937}, increased difficulty in focusing on relevant information, and being more easily distracted by irrelevant information \cite{boksem_effects_2005}.  These patterns closely mirror the challenges participants described as \rev{``blindness effects'' during fatigue, suggesting a possible explanation for the experience.} 

\subsection{Developer Experience} 

Since our goal is to model the developer’s \rev{cognitive} experience during troubleshooting, \change{this section reviews existing definitions and frameworks for Developer Experience, alongside research on the emotional dimensions of software development. We then clarify how these perspectives inform the orientation and goals of our Theory of Troubleshooting.}

\subsubsection{Defining Developer Experience}

Research in Developer Experience shifts the focus from \change{what happens in the code} to understanding the experiences of software developers that interact with the code. \change{We take a comprehensive view of that experience, encompassing the developer’s thoughts, feelings, motivations, and intentions, as well as the situational context: the task at hand, the social environment, the state of the code, and the supporting infrastructure.}

\change{Fagerholm and M\"{u}nch define Developer Experience using the “trilogy of the mind” from social psychology: the interplay of cognitive (attention, memory, problem-solving, decision-making), affective (feeling, emotion), and conative (desire, volition, striving) dimensions \cite{fagerholm_developer_2012}.} Greiler et al. \change{extend this definition and propose} an actionable framework for improvement (DX Framework), defining Developer Experience as how developers think about, feel about, and value their work \cite{greiler_actionable_2023}. \change{The DX Framework identifies actionable factors} that influence Developer Experience, such as development and release, product management, collaboration and culture, and developer flow and fulfillment.  The DX Framework also identifies barriers to improvement and coping strategies developers use when they are unable to improve their experience, such as reducing engagement, no longer speaking up, or leaving their job as a last resort, if they are unable to improve the working conditions \cite{greiler_actionable_2023}.

\change{Fagerholm and M\"{u}nch also distinguish between the verb \textit{experiencing}—a person's stream of perceptions, interpretations, and resulting emotions that occur during interaction—and the noun \textit{experience},} a person's encounter with the system that has a beginning and an end \cite{fagerholm_developer_2012}. While the DX Framework emphasizes the latter, developers often speak about Developer Experience in terms of the former—what they are experiencing in the moment, especially friction, frustration, or flow. \rev{Our models center this verb-oriented form of experiencing—foregrounding the dynamics of thinking, feeling, and striving that unfold during the troubleshooting process.}

\subsubsection{Emotional Dynamics of Developer Experience}

\rev{Emotions are a critical component of the developer's experience that are greatly affected by the circumstances of coding, like getting stuck.} Wrobel found that frustration was the most commonly reported negative emotion in a survey of 49 developers, with 95.9\% having experienced it while coding \cite{wrobel_emotions_2013}. General psychology research by Burleson and Picard shows that the state of being stuck and making no progress is commonly associated with negative emotions like frustration, cognitive fatigue, and distress, while the state of being in flow and making lots of progress is commonly associated with positive emotions \cite{burleson_affective_2004}.  Table \ref{tab:flow_vs_stuck} shows Burleson and Picard's comparison of the experience of being in flow versus being stuck. \change{They also emphasize} that the ability to persevere and sustain motivation through the challenge of being stuck is essential to gaining expertise, and that it is not wise to simply try to eliminate the experience of difficulty \cite{burleson_affective_2004}.  

\begin{table*}[h]
    \renewcommand{\arraystretch}{1.5}
    \centering
    \begin{adjustbox}{max width=\textwidth}
    \begin{tabular}{|p{8.4cm}|p{8.4cm}|}
    \hline
        \textbf{Flow: Optimal Experience} & \textbf{Stuck: Non-Optimal Experience}\\
            \hline
        All encompassing & All encompassing\\
            \hline
        A feeling of being in control & A feeling of being out of control\\
            \hline
        Concentration and highly focused attention & A lack of concentration and inability to maintain attention\\
            \hline
        Mental enjoyment of the activity for its own sake & Mental fatigue and distress caused by engagement with the activity \\
            \hline
        A distorted sense of time & A negative distorted sense of time, taking forever or never ending and taxing endurance (Waybrew, 1984; Czerwinski et al. 2001) \\
            \hline
        A match between the challenge at hand and one's skills & A perceived mismatch between the challenges and one's skill \\
            \hline
        Frequently associated with positive affect & Frequently associated with negative affect \\
            \hline
    \end{tabular}
    \end{adjustbox}
    \caption{Comparison of elements of the Flow Experience versus the Stuck Experience from Burleson and Picard \cite{burleson_affective_2004}}
    \label{tab:flow_vs_stuck}
\end{table*}

\rev{These affective contrasts between flow and stuckness articulated in general psychological terms are echoed in developer‑specific studies.}  M\"{u}ller and Fritz explored these emotional dynamics further in a lab study where 17 developers wore biometric sensors while coding and periodically assessed their emotions and perceived progress \cite{muller_stuck_2015}. They observed a wide range of emotional states, both in valence and arousal. Developers frequently commented on emotions and progress together, noting for example that a lack of progress triggered annoyance. These comments suggest a strong connection between affective state and perceived progress.

\change{M\"{u}ller and Fritz also analyzed developers’ explanations for emotional shifts and found recurring patterns \cite{muller_stuck_2015}.} For example, when the developers struggled to locate or understand the relevant parts of the code as a starting point, this led to negative emotions and feeling stuck.  In contrast, being able to identify the relevant code made the developers feel better, and like they would make progress soon. The developers also distinguished between feeling ``completely stuck'' and feeling ``less stuck,'', and most reported less than maximum flow.  \change{These findings indicate that stuckness and flow are not binary but exist on a continuum.}

\rev{A key finding from M\"{u}ller and Fritz’s study is that one of the biggest shifts in developers’ emotions and perceived progress occurred when writing code and seeing it run \cite{muller_stuck_2015}.} It was not so important whether the output from the code was correct as much as whether the developer succeeded at creating some visible output, which resulted in positive emotions and a perception of progress. \rev{When developers write code, run it, and observe the resulting behavior, they are establishing a feedback loop. The increase in positive emotions occurred when the developer could act on the system and see it respond. While Müller and Fritz document this affective shift, the broader process context that makes this moment emotionally significant remains implicit.  Understanding why establishing a feedback loop changes how the work feels requires a model that situates emotion within the unfolding activity itself, integrating feeling, sense-making, and action over time. }

\subsubsection{Toward Understanding the Developer's Cognitive Experience}

\rev{For many developers, the internal experience of developing software is largely tacit knowledge that is unavailable to conscious inspection.  The developer might have an intuition about what to do, but be unable to describe their process.  As identified by Perscheid, many developers found it difficult to explain their approach to debugging and resorted to demonstrating, rather than explaining \cite {perscheid_studying_2017}.  After many years of experience, however, developers often gain self-awareness and can describe their internal processes more clearly.  Our study provides qualitative accounts from many experienced developers, taking advantage of this self-awareness, allowing us to construct models of the developers' internal experiences and make them explicit.} \rev{We use the term \textit{cognitive experience} to refer to the developer’s internal experience of thinking, feeling, and striving over time as an integrated whole.}

\section{Methodology} \label{methodology}

To design the study, we followed Charmaz's philosophy and approach to Constructivist Grounded Theory \cite{charmaz_constructing_2014} to develop an empirically-grounded \rev{theory}.  Constructivist Grounded Theory (CGT) provides an iterative method of data collection and theoretical sampling to achieve theoretical saturation, and a data coding and analysis process that leverages initial coding with gerunds, constant comparison, and memo writing, resulting in emergent theory. \rev{While we did our best to closely follow CGT’s analytic procedures, where we diverge, we clearly explain in the remainder of this section.}  The study was carried out with the approval of the Ethics Committee at University of Victoria, and with the informed consent of participants (Ethics Certificate \#23-0106).

The primary question in a grounded theory study is, ``what is happening in the scene?'' and the researcher interprets what is happening, and through their codes, makes the relationships between implicit processes and structures visible \cite{charmaz_constructing_2014}.  The constructivist philosophy starts with the assumption that social reality is a construction, and rather than assuming that the researcher is an objective, neutral observer, we take the researcher's perspective, privileges, and interactions into account as an inherent part of the research context \cite{charmaz_constructing_2014}.  In recognizing how the researcher's preconceptions may shape the analysis, we work to mitigate the risks of researcher bias in designing the interviews for the study, as we highlight in sections \ref{interviews} and \ref{followups}.

In this section, we review the methods used in the study for recruiting, interviewing, and data analysis.  We start with the perspective of the researchers in Section \ref{perspective} to give context for the decisions and preconceptions that may shape the analysis.  Next, we review the recruiting process in Section \ref{recruiting}, the interview process in Section \ref{interviews}, the data analysis in Section \ref{data-analysis}, and finally the process of evaluating resonance and getting feedback in Section \ref{followups}.

\subsection{Perspective of the Researchers} \label{perspective}

Charmaz discusses the influence of the researcher when conducting interviews, and highlights that the way participants identify the researcher influences what they will share \cite{charmaz_constructing_2014}. \change{The interviews and data analysis were all done by the first author as part of a PhD dissertation. The second author met with the first author regularly to review and discuss the emerging theoretical categories and data analysis strategy.}

The first author comes with over 20 years experience as a software engineer, is an industry conference and keynote speaker \cite{vmware_tanzu_arty_2021}, and author of the book, \change{Idea Flow \cite{starr_idea_2012}.}  This position gave us the opportunity to build a peer-level rapport with developers as a fellow software engineer, and construct models that reflect an ``insider's view'' of the world of developers.  In this influential position, we were also able to recruit many senior software engineers with substantial experience and wisdom to participate in the study.

\change{The first author's book, Idea Flow, discusses data collection and models related to flow and troubleshooting, leading to potential biases that needed to be mitigated through careful design of the study.} The first author's perspective has also been shaped as a research assistant and PhD student at CHISEL (Computer Human Interaction and Software Engineering Lab) at University of Victoria, while collaborating with other researchers on a variety of projects.  The author's research interests span software engineering, cognitive science, and HCI.

\change{The second author is a research professor at the University of Victoria studying human-centered software engineering with a long-standing interest in how developers understand, maintain, and evolve complex software systems. Her perspective is shaped by two decades of conducting empirical and collaborative research with software professionals across academia and industry. In this study, she contributed to study design and interpretation of the findings. She brings a lens informed by developer experience, socio-technical systems, and a commitment to understanding the cognitive and collaborative challenges developers face when troubleshooting. Her ongoing collaborations with industry partners have deepened her insights into the practical realities of developer work, especially in the context of modern tooling and AI-assisted workflows.}

\rev{Many of the data analysis decisions were decided by the first author, with input from the second author.  Different researchers reviewing the same interviews in a different sequence, noticing a different set of patterns, would have likely resulted in a different theory.  Constructivist Grounded Theory acknowledges that all grounded theories are inherently interpretive, and has these limitations. In conducting the research, we adopted a posture of ``strong beliefs, loosely held'' and implemented protocols such as preparing semi-structured interview questions to reduce the researcher’s influence on participant responses. The follow-up interviews and participant feedback also provided an additional layer of validation and iterative refinement of our theory.  While these protocols mitigate bias, they may not fully eliminate it.} 

\rev{While CGT process does not call for agreement sessions commonly used in some qualitative traditions, instead a deep solo engagement is typical in Constructivist Grounded Theory, where the researcher's interpretive role, sustained immersion, and reflexive memoing are central to theory development.  We designed our study to construct grounded theory and minimize researcher bias, to see if a connection between troubleshooting and cognitive fatigue would emerge from the data on its own.}

\subsection{Recruiting Participants} \label{recruiting}

\rev{We recruited study participants using convenience sampling.}  \change{We created a signup page describing the study and 60-minute remote Zoom interview, and invited developers to participate in a post on Mastodon federated social media network.} \rev{11 developers signed up for the study from the social media post, and eight of those followed through with the interview.}

To increase the diversity of developer participants, we also recruited participants using search on LinkedIn.  We focused our LinkedIn search on developers with five or more years of experience primarily within the United States or Canada with some in Europe, striving for balance across front-end, backend, and tools developers, and ensuring representation of women and non-binary developers.  \rev{We sent} 113 developers LinkedIn InMail messages about the study, 29 developers replied 'yes', and 19 of those developers followed through with the interview.  This yielded 27 developer participants representing 570 total years of experience, \rev{ranging from five to 42 years, with most participants having over a decade of experience.}  \change{Table \ref{tab:participant-demographics} shows a summary of the demographics of the participants, with a more detailed table available in the Appendix.  Additionally (not \rev{shown} in the table), 15 participants were from the US, eight from Canada, three from Europe, and one unreported.} \change{ No compensation was offered to participants.}

\begin{table}[h]
\centering
\renewcommand{\arraystretch}{1.3} % slightly tighter vertical space
\setlength{\tabcolsep}{4.5pt}     % slightly tighter horizontal 
\small
\begin{tabular}{|l|c|l|c|l|c|l|c|l|c|l|c|}
\hline
\textbf{Experience} & \textbf{N} &
\textbf{Education} & \textbf{N} &
\textbf{Company Size} & \textbf{N} &
\textbf{Team Size} & \textbf{N} &
\textbf{Gender*} & \textbf{N} &
\textbf{Work Env.} & \textbf{N} \\
\hline
Over 20 yrs & 15 & Post-graduate & 8 & > 1000 people & 12 & > 10 people & 4 & Woman & 8 & Remote & 24 \\
13–19 yrs   & 6  & Bachelor’s & 17 & 51–1000 people & 9 & 6–10 people & 12 & Man   & 18 & Co-located & 1 \\
8-12 yrs    & 5  & Some college & 1 & < 50 people & 6 & <= 5 people & 11 & Non-binary & 1 & Hybrid & 2 \\
<= 7 yrs    & 1  & Unreported & 1 & & & & & & & & \\
\hline
\end{tabular}
\caption{Summary of participant demographics (N = 27). *Gender based on LinkedIn profiles; not self-reported.}
\label{tab:participant-demographics}
\end{table}

Due to low LinkedIn response, \rev{we had to send many interview requests, and conducted} the initial 27 interviews before formal data analysis \rev{to proceed efficiently and not keep participants waiting}.  While this diverged from the ideal CGT protocol of interleaving data collection and analysis, we consulted an experienced CGT researcher to review our strategy and ensure theoretical saturation, \rev{as noted in the acknowledgments}. Although CGT encourages theoretical sampling to drive recruitment, we found this was not feasible given \rev{our focus on understanding the developer's internal experiences, which were not identifiable} through external recruitment criteria. Instead, we structured our interviews around a consistent prompt asking participants to walk through a recent troubleshooting experience. This approach enabled us to capture diverse instantiations of the phenomena of interest across a wide range of developer contexts. To mitigate the risk of \change{incomplete theoretical development} because of only a single round of interviews, we conducted an additional seven follow-up interviews within the same set of developers to get feedback, evaluate resonance, \change{and iterate on the models until we reached theoretical saturation.}

\subsection{Developer Interview Process} \label{interviews}

\rev{For our main round of 27 interviews}, we designed a semi-structured, 60-minute interview protocol with open-ended questions on individual and collaborative troubleshooting experiences. All interviews were conducted by the first author over Zoom. \change{To mitigate researcher bias, we used a prepared question set to reduce the likelihood of leading participants.}  For example, although we \rev{speculated} a relationship between troubleshooting and cognitive fatigue, we did not ask participants directly about cognitive fatigue. Instead, we asked developers how they felt during troubleshooting, allowing relevant experiences to emerge naturally. While the interview generally followed the prepared questions, we occasionally asked clarifying follow-ups, taking care to avoid suggestive phrasing.

The questions followed a three-part structure, where we \rev{asked the developers to} 1) describe their environment/context, 2) reflect on general experiences of troubleshooting, and 3) reflect on a specific experience of troubleshooting.  We followed a similar protocol for related topics of collaborating on troubleshooting and asking for help when stuck.  The full interview protocol is provided in the Appendix. Our first \rev{interview} revealed a need to reword some questions for clarity and reorganize them for better flow. The second interview went smoothly, and we used the revised questions for the remaining 25 participants. Since the CGT method does not require structured interviews, all interviews were included with the findings.

At the beginning of the interview, we informed the participants that it was \rev{acceptable} to speak in ``familiar developer terms'' because the interviewer had substantial experience as a software engineer.  \change{The interviewer’s software development background allowed participants to share more directly, without needing to translate for someone outside their world. This shared experience helped build rapport, leading to detailed responses and rich qualitative data.}  At the end of \rev{each} interview, participants were asked if they would be open to participate \change{in follow-up research,} and all 27 developers agreed to be contacted. \change{The follow-up interviews are discussed in Section \ref{followups}.}

\subsection{Data Analysis} \label{data-analysis}

\begin{figure*}[h]
\centerline{\includegraphics[scale=0.41]{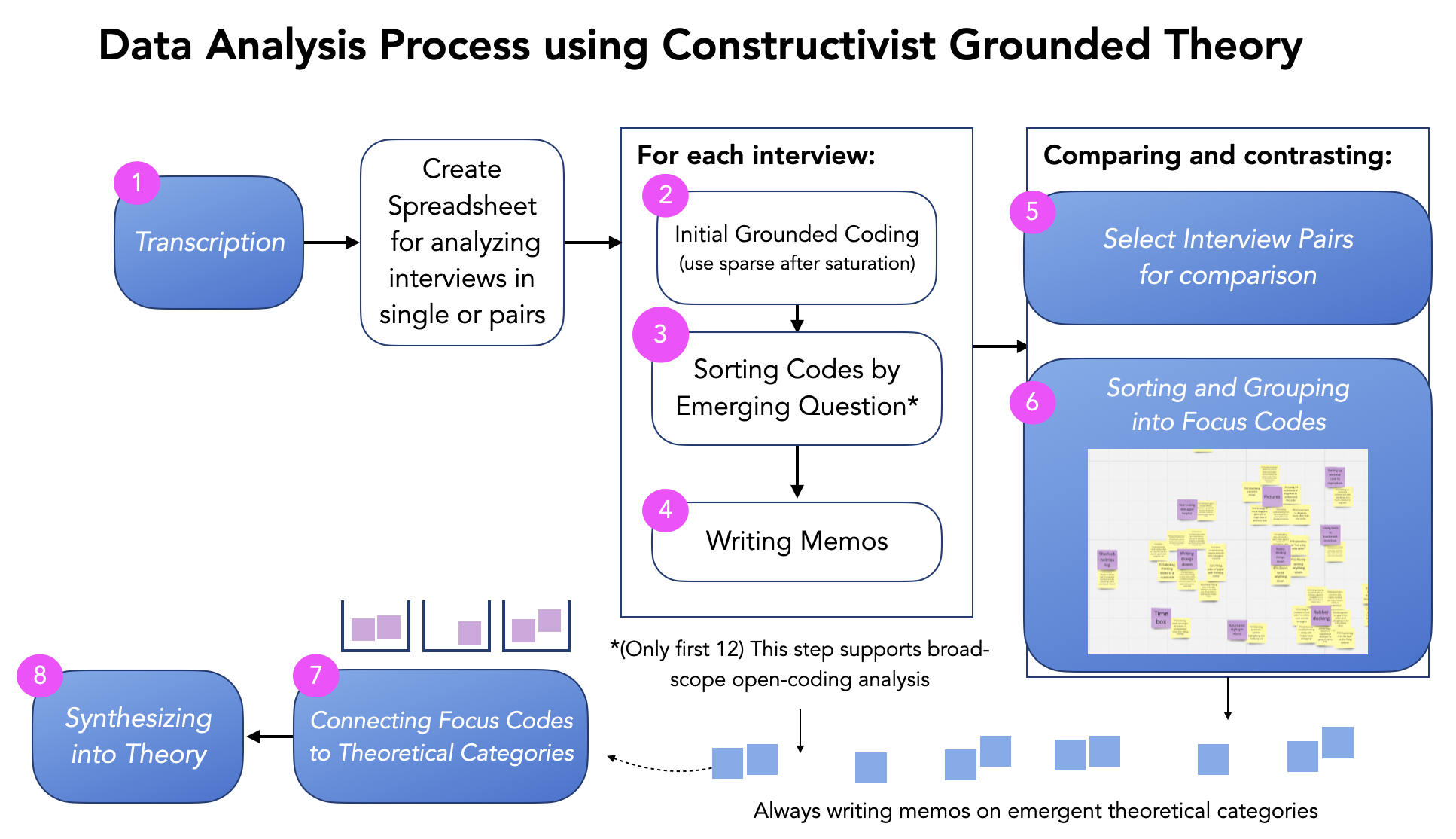}}
\caption{Overview of the data analysis process used in applying Constructivist Grounded Theory}
\label{fig:data_analysis}
\end{figure*}

\change{We followed CGT methods from Charmaz \cite{charmaz_constructing_2014}, beginning with line-by-line initial grounded coding, constant comparison, and iterative memo writing to develop theoretical categories.} In this section, we walk through the steps of our data analysis process, \change{as illustrated} in Figure~\ref{fig:data_analysis}, and provide examples to demonstrate how we applied the CGT research method to analyze the interview data.

\subsubsection{Transcription}

We recorded the Zoom interviews on video, and transcribed the audio locally using MacWhisper 4, which leverages OpenAI's transcription technology, Whisper, with the Medium (English) model. As we watched the videos of the interviews, we fixed any errors we found in the transcripts. The output from the transcription process (Step 1 in Figure \ref{fig:data_analysis}) is a line-by-line transcript that we loaded into a spreadsheet, with additional columns to \rev{track the  initial grounded codes.} 

\subsubsection{Initial Grounded Coding}

\begin{figure*}[h]
\centering
\includegraphics[width=\linewidth]{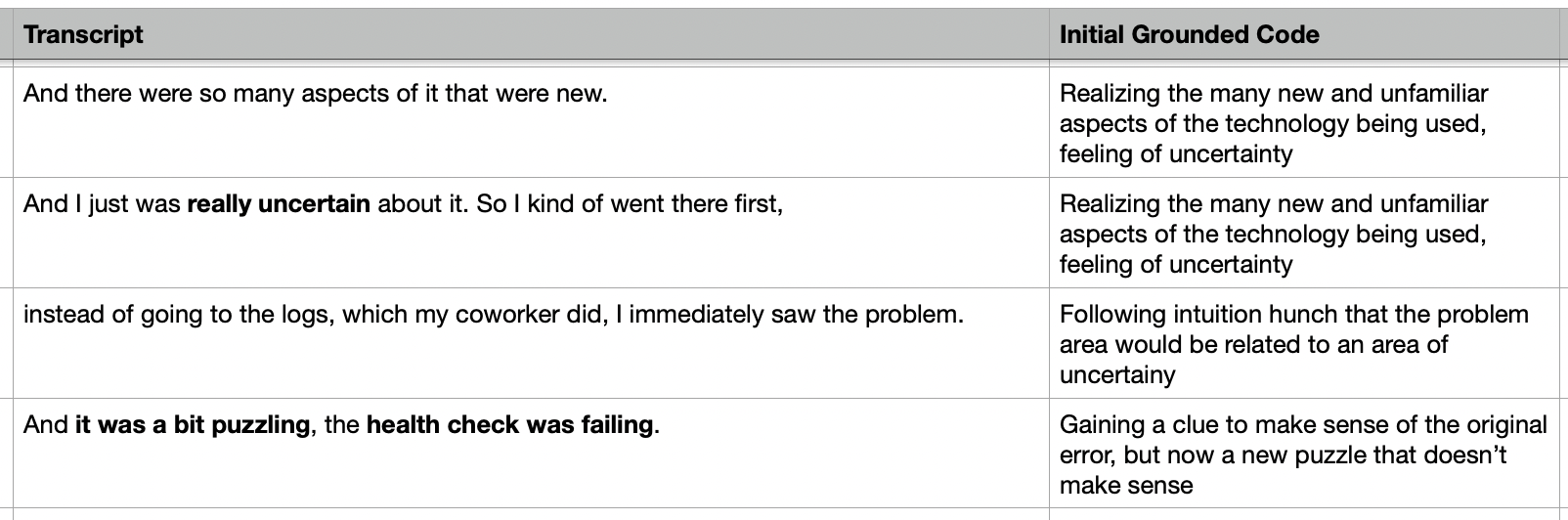}
\caption{Example of the initial grounded coding from P25}
\label{fig:grounded_coding}
\end{figure*}

To create the initial grounded codes (Step 2 in Figure \ref{fig:data_analysis}), we watched the video recording while reading through the transcript, then bolded key phrases that referenced important concepts or points being made by the developer.  Bolding phrases in the transcript aided in comprehension and made the transcript easier to scan and review multiple times.  We then coded line-by-line with \rev{high-resolution} descriptive codes of what was happening, starting with a gerund, an -ing suffix word, a technique suggested by Charmaz to help the researcher focus on the actions and processes \cite{charmaz_constructing_2014}.  Figure \ref{fig:grounded_coding} shows an example of how we created initial grounded codes starting with gerunds.

In places where gerunds felt awkward to use, we allowed for alternatives of starting with active verbs (-s suffix words). If we wanted to describe the situation, or anything other than an action or process, we abandoned these rules and aimed to be pragmatic with designing codes that made sense for the context.  We allowed for multiple initial grounded codes per line using multiple spreadsheet columns, to help capture the complexity of what was happening.

\rev{Because the troubleshooting process involves substantial contextual nuance and our aim was to understand the micro-level shifts in the developer’s internal cognitive experience, we intentionally used high-resolution descriptive codes to preserve fidelity to the developer’s lived experience. Unlike shorter codes more typical in early stages of grounded theory, our initial grounded codes were often longer and more detailed, functioning as atomic units of meaning that could be understood independently of the original transcript context. While some initial grounded codes were common, such as ``Using printf to debug,'' the majority of the codes were unique, yielding 1032 such codes in analyzing the first 12 interviews.  This high-resolution approach enabled fine-grained comparison during the constant comparison phase to support theory development.}  Later in the process, we sorted and grouped similar initial grounded codes into focus codes to raise the level of abstraction.

\subsubsection{Sorting Codes by Emerging Questions}

\rev{As we began reviewing initial grounded codes from the interviews, we identified a list of 16 emerging questions arising from the data.  For each interview, we created another spreadsheet tab and sorted the initial grounded codes by the emerging question they helped answer (Step 3 in Figure \ref{fig:data_analysis}).  As new questions arose during the analysis, we added questions to the list, then re-analyzed prior interviews as needed to ensure consistency of the analysis across interviews.}  Sorting the initial grounded codes \change{by emerging question} gave us a way to review related sets of initial grounded codes \rev{and sort them on Miro boards (Section \ref{sorting}), allowing us to more easily make comparisons and discover insights.}

The emerging questions explored three main areas of inquiry: Developer Context, Troubleshooting, and Asking for Help. If an initial grounded code was not helpful in answering any of the emerging questions, or otherwise an important insight, it was filtered from further analysis. The full set of emerging questions are available in the Appendix. 

\subsubsection{Writing Memos for Emerging Categories}

As theoretical categories emerged \rev{throughout the analysis process}, we created memos for each theoretical category noting relevant insights from the developers along with our thoughts (Step 4 in Figure \ref{fig:data_analysis}).  \rev{We created these memos while listening to interviews, during the initial grounded coding process, as well as during constant comparison—the moment a key concept emerged.  We continued to add notes to our theoretical category memos whenever the same concept emerged in subsequent interviews, helping us to make connections of different perspectives across developers.}

\subsubsection{Selecting and Comparing Interviews} \label{pairs}

\rev{In deciding the sequence of interviews to review and analyze, we started by selecting two interviews at a time (Step 5 in Figure \ref{fig:data_analysis}), choosing pairs of interviews to compare and contrast. The constant comparison process involves comparing many different types of data—most typically by comparing emerging codes, and sorting and grouping them to identify patterns \cite{charmaz_constructing_2014}.  Charmaz also highlights the value of comparing interviews with interviews, a strategy we extended by deliberately selecting and comparing \textit{a pair} of interviews.  We performed the initial grounded coding and sorting codes by emerging question for both interviews, then followed with writing a 
``compare and contrast'' memo to discuss the similarities and differences.}

\rev{When choosing the pairs, we identified interviews that had interesting comparisons of similarities and differences.} For example, the first pair of developers we compared, \rev{both developers} worked in a web cloud infrastructure environment with microservices.  One developer's environment had a significant investment in automation and observability tooling and a healthy culture.  The other developer worked in a legacy microservices environment that was spinning out of control with compounding difficulties.  By selecting pairs of interviews with clear similarities, many distinctive patterns \rev{also} stand out in the contrast, leading to a rapid discovery of new theoretical insights.  We found this pairwise interview comparison technique to be quite fruitful. 

To ensure we considered a broad range of perspectives, we prioritized interviews that \rev{covered a variety of developer roles, including backend developers, full-stack developers, technical leads, tooling developers, and consultative coaches}.  \change{We acknowledge the subjectivity of these decisions is an inherent limitation with CGT.} We reviewed 12 interviews before reaching theoretical saturation:

\begin{itemize}
\item {\textbf{P25 and P11}: Two developers both working in a web cloud infrastructure environment with microservices, one environment healthy, another full of challenges}
\item {\textbf{P06 and P19}: Two developers both working primarily in backend development, one working in a more legacy environment, the other a more modern environment}
\item {\textbf{P02 and P14}: Two developers both working in dev experience, dev tooling, dev relations space, building tools that made developers lives easier.}
\item {\textbf{P16 and P10}: Two developers working in team lead, technical leadership roles in mature environments leveraging sophisticated automated tooling}
\item {\textbf{P04 and P23}: Two developers working in consultative/coaching type roles with deep expertise in being able to teach, coach, lead in technical practices}
\item {\textbf{P15 and P27}: Two developers working in full-stack development with experience in architecture and team dynamics}
\end{itemize}

\rev{Once we reached theoretical saturation, we adjusted our process for reviewing the remaining interviews, narrowing the theoretical scope under consideration and reviewing one interview at a time rather than in pairs. We describe our adapted process for reviewing the remaining interviews in Section \ref{sparse}.}

\subsubsection{Sorting and Grouping into Focus Codes} \label{sorting}

\rev{To facilitate deeper analysis}, the spreadsheet data, including interview transcript and initial grounded codes, were loaded into a Postgres database.  We created an automated data pipeline to load and transform the data to produce reports.  We used Gradle, Docker, Liquibase, and Springboot to create the data pipeline with an existing boiler plate configuration from another software project.  The data pipeline allowed us to automate the process of generating reports, and develop simple tools to extract quotes from the transcripts.  \rev{We used the database to analyze the data directly with queries, and iteratively re-generate reports as we made progress with coding.}

\rev{The initial reports included a set of 16 emerging question reports with initial grounded codes and corresponding participant number sorted by emerging question.}  \rev{We created a Miro board} for each emerging question, then imported the table rows of the report into Miro as sticky notes (Step 6 in Figure \ref{fig:data_analysis}).  We then sorted and grouped the sticky notes into focus codes that represented the common idea or concept.  Once all the stickies were sorted, we reviewed each focus code group to see if the grouping needed to be sorted further into multiple groups, and revised the focus codes and groups as needed.  We continued this process of sorting and grouping until all focus codes were clearly named, and the initial grounded codes within each group had good cohesion.  Links to the 16 Miro boards, and a data archive that includes all 1032 grounded codes can be found in the Appendix.

\begin{figure}[h]
\centering
\includegraphics[scale=0.47]{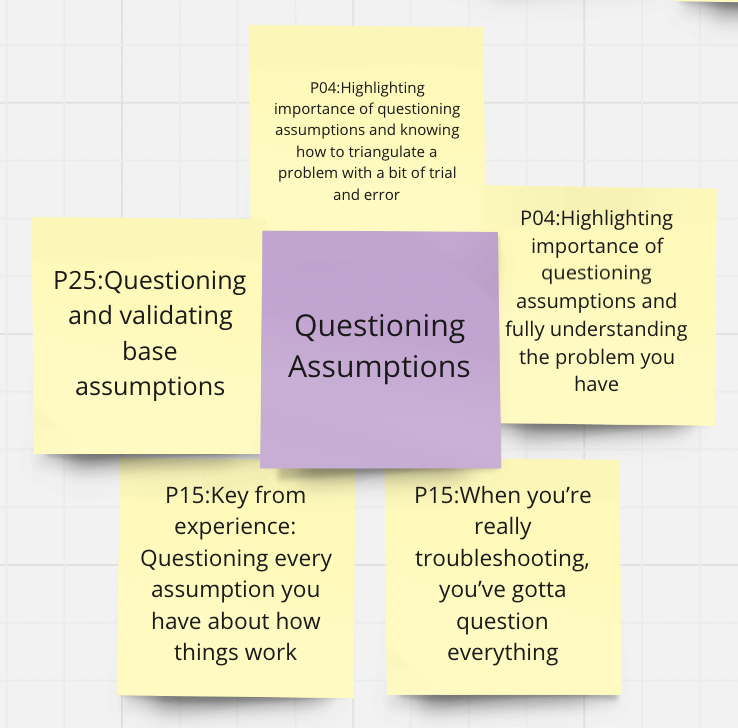}
\caption{Example cluster of stickies from the ``Troubleshooting Strategies'' Miro board with the ``Questioning Assumptions'' focus code sticky surrounded by initial grounded code stickies with corresponding participant numbers that indicate the source of the data.}
\label{fig:focus_cluster}
\end{figure}

For an example of what this process looks like, the ``Questioning Assumptions'' focus code in Figure \ref{fig:focus_cluster} is represented by a purple sticky in Miro surrounded by five yellow stickies representing each related initial grounded code.  The stickies also include the participant number where the initial grounded code originated, and shows how the ``Questioning Assumptions'' construct is grounded in data from three different developers, P04, P15, and P25.

The \rev{analysis of the first 12 interviews} resulted in 1032 initial grounded codes sorted across 16 Miro boards by emerging question.  Through this sorting and grouping process, the initial grounded codes were consolidated into 451 focus codes, 45 related to developer context, 231 related to troubleshooting, and 161 related to collaboration.  \rev{After this broad analysis of the first 12 interviews and our theoretical area of focus being well-supported in the data, we determined we had reached theoretical saturation, and narrowed our area of focus for the remainder of the analysis.}

\subsubsection{Narrowing our Area of Focus }

\rev{Our initial analysis surfaced 73 theoretical categories which we grouped by primary orientation into four broad themes: Experience, Strategy, Collaboration, and Thriving. As is common in Constructivist Grounded Theory (CGT) studies \cite{charmaz_constructing_2014}, the early stages of analysis cast a wide net to map the breadth of meaningful patterns present in the data. From this broad theoretical foundation, we narrowed our focus to the core phenomenon of the confusion experience and the cognitive processes involved in overcoming the confusion. Our theory centers on eight theoretical categories that most directly capture the developer's \rev{cognitive} experience of troubleshooting:}

\vspace{1em}

\begin{minipage}[t]{0.48\textwidth}

\begin{itemize}
\item {Confusion Experience}
\item {Trouble in the Creation Process}
\item {Trying to Gain Clarity}
\item {Poking and Seeing}
\end{itemize}
\end{minipage}
\hfill % Adds horizontal space between minipages
\begin{minipage}[t]{0.48\textwidth}
\begin{itemize}
\item {Elucidating the Problem}
\item {Frustration vs Confidence}
\item {Experiential Intuition}
\item {Figuring It Out}
\end{itemize}
\end{minipage}

\vspace{1em}

\rev{The remaining theoretical categories extend beyond the scope of this paper and offer opportunities for future exploration. The full set of 73 theoretical categories is provided in the Appendix.}

\subsubsection{Connecting Focus Codes to Theoretical Categories}

At this point, we had both a bottom-up process that started with initial grounded codes being sorted and grouped into focus codes, as well as a top-down process that started with \rev{theoretical categories emerging during memo writing}. The next step was building connections between the focus codes and the theoretical categories (Step 7 in Figure \ref{fig:data_analysis}).

To make the connections \rev{between the focus codes and theoretical categories}, we created \rev{an additional theoretical category} memo for each of the eight theoretical categories, then looked through the Miro boards to find all relevant focus codes, and copied the focus codes, and all related initial grounded codes with corresponding participant numbers, into the memo.  We also used the database to search across all data in the Miro boards and find the focus codes that corresponded to key words remembered from an interview, to improve the thoroughness with connecting the theoretical categories to the data.  We extracted quotes from the interview transcripts related to each theoretical category, and added these to the memos too.  After collecting all relevant data to each theoretical category memo, we wrote additional notes to explain the theoretical category in terms of the evidence found in the data, and revised the names of the theoretical categories as the meanings became more clear. 

\subsubsection{Sparse Grounded Coding} \label{sparse}

\begin{figure*}[h]
\centering
\includegraphics[width=\linewidth]{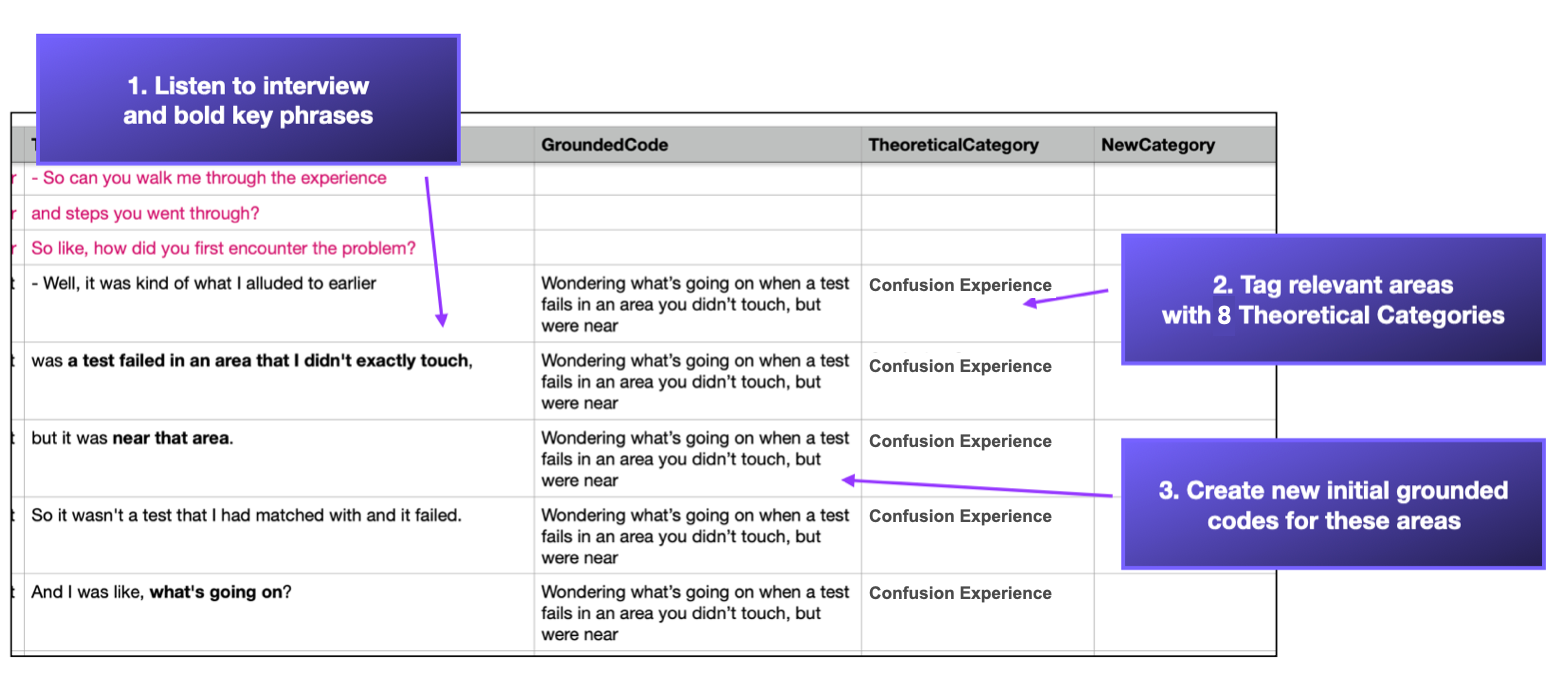}
\caption{Example of the sparse grounded coding process}
\label{fig:sparse_grounded_coding}
\end{figure*}

For the remaining 15 interviews, we used a sparse grounded coding process: we reviewed each interview individually (with no pairing), \rev{identified segments relevant to the eight core theoretical categories, then applied line-by-line initial grounded coding only to those segments.} \change{Figure~\ref{fig:sparse_grounded_coding} illustrates an example of this process.}

While these remaining 15 interviews did not yield new theoretical categories \rev{within} our core theory, they contributed 309 new initial grounded codes \rev{that added nuance and contrast within each theoretical category.}  We continued to load the transcript data encoded with theoretical categories and initial grounded codes into our database to search for keywords, examples, and extract quotes.  \rev{We also added a new report for each of the core theoretical categories that included the initial grounded codes with corresponding participant numbers for all 27 interviews.} \rev{These additional reports allowed us to continue comparing codes during the writing process, and are available in the Appendix. }

\subsubsection{Reaching Theoretical Saturation} \label{saturation}

From the 27 developer interviews, we reached theoretical saturation after analyzing the first 12, \rev{meaning} no new theoretical categories were emerging within our core area of focus. Theoretical saturation was assessed based on the recurrence and completeness of our key theoretical categories—such as the confusion experience and experiential intuition—which appeared consistently across nearly all interviews.

\rev{To support transparency of the breadth and depth of theoretical grounding, the Appendix includes the eight theoretical category reports comprising 681 grounded codes across all 27 interviews, a summary table showing participant counts and initial grounded codes for each theoretical category, and a more detailed breakdown of the counts by participant. Most theoretical categories are supported by over 20 participants, reflecting the robustness and saturation of the theory.  The Appendix also includes a diagram of the data analysis terminology we use (e.g. types of codes) and how they relate. }

\subsubsection{Synthesizing into a Grounded Theory}

To synthesize and integrate the theoretical categories into a \rev{grounded theory} (Step 8 in Figure \ref{fig:data_analysis}), we leveraged the writing process \change{as suggested by Charmaz \cite{charmaz_constructing_2014}.}  \rev{We began by explaining our findings with each theoretical category, creating models and drawing diagrams to describe the phenomena, then integrating the models into a cohesive theory. During the process, we considered our research question of what developers were thinking, feeling, and striving for over time.}  Since the theoretical categories were highly interrelated, as we began writing, \rev{we pulled in concepts from other theoretical categories as needed to create a cohesive theory.}

\subsection{Evaluating Resonance and Getting Feedback} \label{followups}

To evaluate \rev{how} our emerging theory resonated with developers \rev{and get feedback}, we conducted seven follow-up interviews over Zoom. \rev{In line with Charmaz’s guidance on reinterviewing with a focus on theoretical categories \cite{charmaz_constructing_2014}, we prioritized participants who had named or richly described the tacit phenomena that gave rise to our theoretical categories.  We also aimed for representation across roles and genders, to ensure a broad range of perspectives.}

\rev{During the follow-up interview, we asked the developers to describe what they saw in each diagram before offering any explanation. Next, we asked whether the model resonated with their experience, what they would change, and a resonance rating on a one to five scale. } \change{This process created space for unprompted reactions and both qualitative and quantitative feedback. We explicitly told participants that disagreement was welcomed and that our goal was to refine the models, helping to create a psychologically safe context for feedback and reduce response bias. This led to a series of diagram iterations and rich discussions about the phenomena.} We continued \rev{an} iterative process of refining the models until we reached a point of two consecutive participants rating the models as highly resonant with no significant changes.  \change{The follow-up interview questions with our initial diagrams can be found in the Appendix, demonstrating our process and how the models evolved.} 

\change{We also emailed a draft of the paper to participants to read and give feedback. The feedback provided additional evidence of resonance, which we discuss in Section \ref{mentors}.} \change{The follow-up interviews and additional written reflections were instrumental in validating the resonance of our theory and ensuring we had reached theoretical saturation.}

\subsection{Limitations}

To recruit the 27 participants in our study, we used a convenience sampling method, which had several limitations. Our \rev{Mastodon social media post reached} developers \change{connected to} the first author's network, which \rev{could introduce bias.} Our LinkedIn search was limited geographically to the United States, Canada, and Europe, reducing cultural representation. Nine of our 27 participants were women or non-binary, and we \rev{included} developers from front-end, back-end, and \change{tool engineering roles}, but we may have missed important perspectives that would have shaped the theory differently.

\change{While our interview-based approach was appropriate for eliciting developers’ internal experiences of troubleshooting, it did not incorporate direct observation of developer behavior during real-time troubleshooting. Future work that complements these internal accounts with observational or screen-recorded studies could verify and deepen understanding of the enacted strategies and contextual dynamics that surround the experiences described in this paper.}

\section{Theory of Troubleshooting} \label{findings}

\begin{figure*}[h]
\centerline{\includegraphics[scale=0.4]{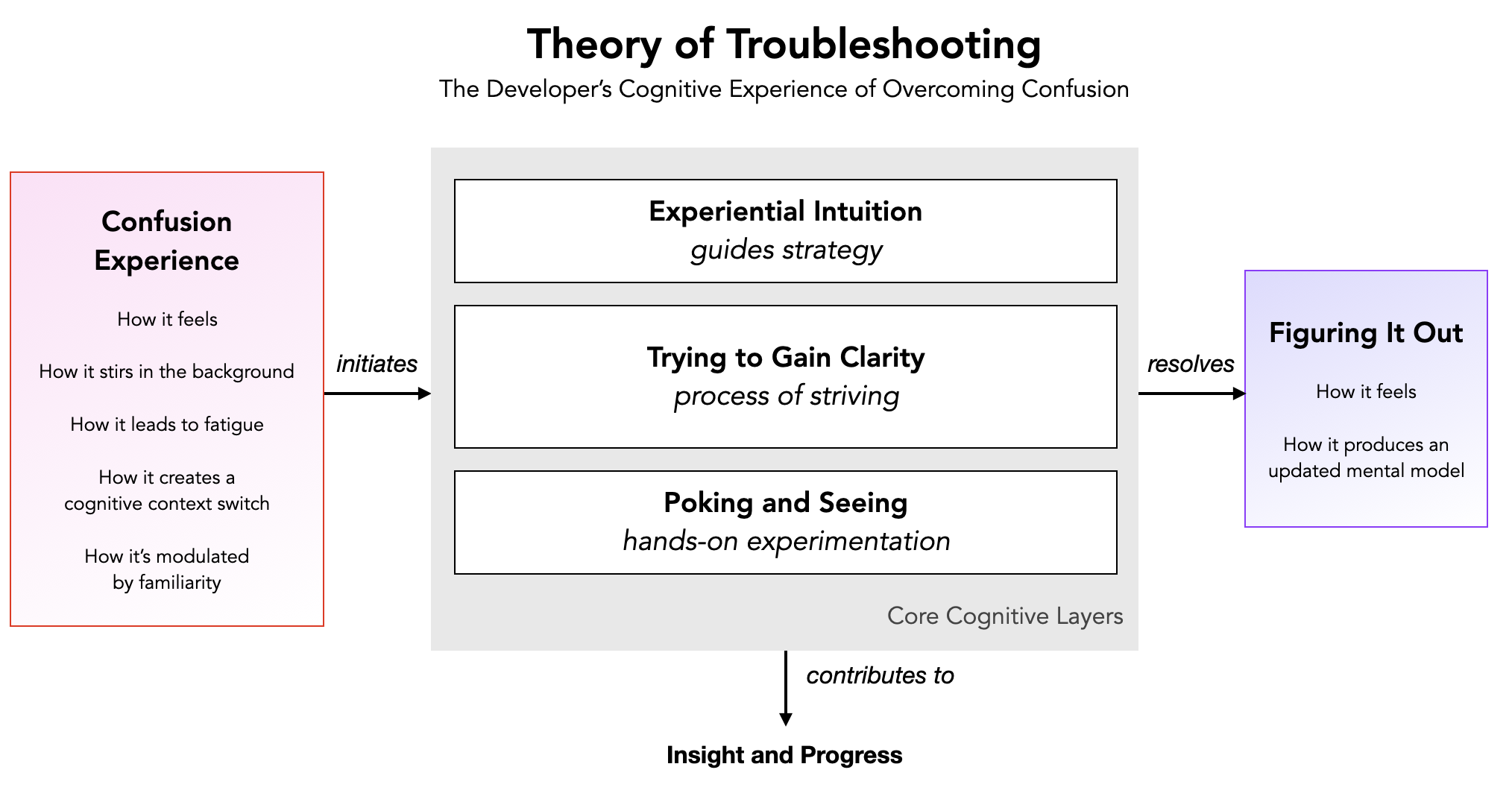}}
   \caption{\rev{The developer’s cognitive experience of overcoming confusion begins with the \textit{confusion experience}, which initiates three interrelated cognitive layers: \textit{experiential intuition} (which guides strategy), \textit{trying to gain clarity} (the process of striving), and \textit{poking and seeing} (hands-on experimentation). These layers contribute to insight and progress. The process ultimately resolves when the developer is \textit{figuring it out.}}}
  \label{fig:theory_of_troubleshooting}
\end{figure*}

In this section, we introduce \rev{the} Theory of Troubleshooting \rev{that followed from our analysis:} \textit{the cognitive problem-solving process of identifying, understanding, and constructing a mental model of the cause of an unexpected system behavior}. \rev{To delineate the scope of our theory,} we begin by describing the different contexts in which troubleshooting occurs.  We then present our Theory of Troubleshooting.  We begin with the \textit{confusion experience}—\rev{the developer's initial encounter} with unexpected system behavior. We then examine the \rev{process of striving} in \textit{trying to gain clarity}, \rev{which involves} the \rev{hands-on experimentation} of \textit{poking and seeing} that drives the learning, and the role of \textit{experiential intuition} in \rev{guiding the strategy.} Finally, we describe the experience of \textit{figuring it out}—when the cause of the unexpected behavior is discovered, \rev{and the developer} overcomes the confusion. Figure \ref{fig:theory_of_troubleshooting} provides an overview of the theory.

\subsection{Contexts of Troubleshooting} \label{when_troubleshoot}

\change{Troubleshooting arises in two primary contexts described by participants: while writing code and building new features (what we refer to as the \textit{creation process}), and while investigating a bug report or deployed system issue. In both cases, developers described shifting into a distinct cognitive mode recognizable as troubleshooting.} \change{During the creation process, developers might run a test or execute the application and encounter unexpected behavior. Participants described these moments as noticeable shifts in cognitive mode. As P23 noted:}

\begin{quote}
\textit{``You're writing code, you're building something, and something's wrong. And this happens hundreds of times a day… But you kind of drop into troubleshooting mode for a moment here and there.''}
\end{quote}

\change{P25 offered a similar reflection, emphasizing how troubleshooting is embedded in the flow of building:}

\begin{quote}
\textit{``There's sort of like, little localized troubleshooting, that is very much a part of the creation process, where I'm testing the boundaries or seeing how something should fit together… and it's part of the building process.''}
\end{quote}

\change{These examples of troubleshooting during the creation process resemble what Alaboudi and LaToza describe as “debugging episodes” \cite{alaboudi_what_2023}.  In both cases, the developer is initially focused on a change task, such as adding a new feature.} \change{When they run the code and encounter unexpected output, the developer realizes their mental model is wrong or insufficient somehow—what we term a \textit{puzzling event}. This moment triggers a shift into troubleshooting, where the developer must reconstruct their mental model to account for what is happening. In this context, troubleshooting is often invisible, perceived as just part of programming.}

\change{The second context occurs when developers begin with a bug report, often involving a deployed system. In these situations, troubleshooting is initiated when the developer begins trying to understand the cause of the bug. The process may span a wide range of tools and layers of infrastructure. P25 contrasted these two contexts in the setting of a small company with a cloud-based microservices architecture:}

\begin{quote}
\textit{``There's the small ones while you're working, or while I'm building a feature, versus a problem that's in production that's weeks later, or months later, or maybe you've just done a deploy and there's a problem. Those are different beasts, because now you're looking at many more systems. I might be troubleshooting something that I didn't work on, that's interfacing with someone else's code, or a latent bug in the sense that maybe something wasn't prepared for a type of load that's now occurring. So in local cases, I'm going to do a lot more running, a lot more local debugging, and then in the web system I'm going to be doing a lot more infrastructure approaches where I'm looking at logs and I'm looking at graphs and looking at the holistic system, especially when I'm looking at a system that has many servers.''}
\end{quote}

\begin{figure*}[h]
\centerline{\includegraphics[scale=0.45]{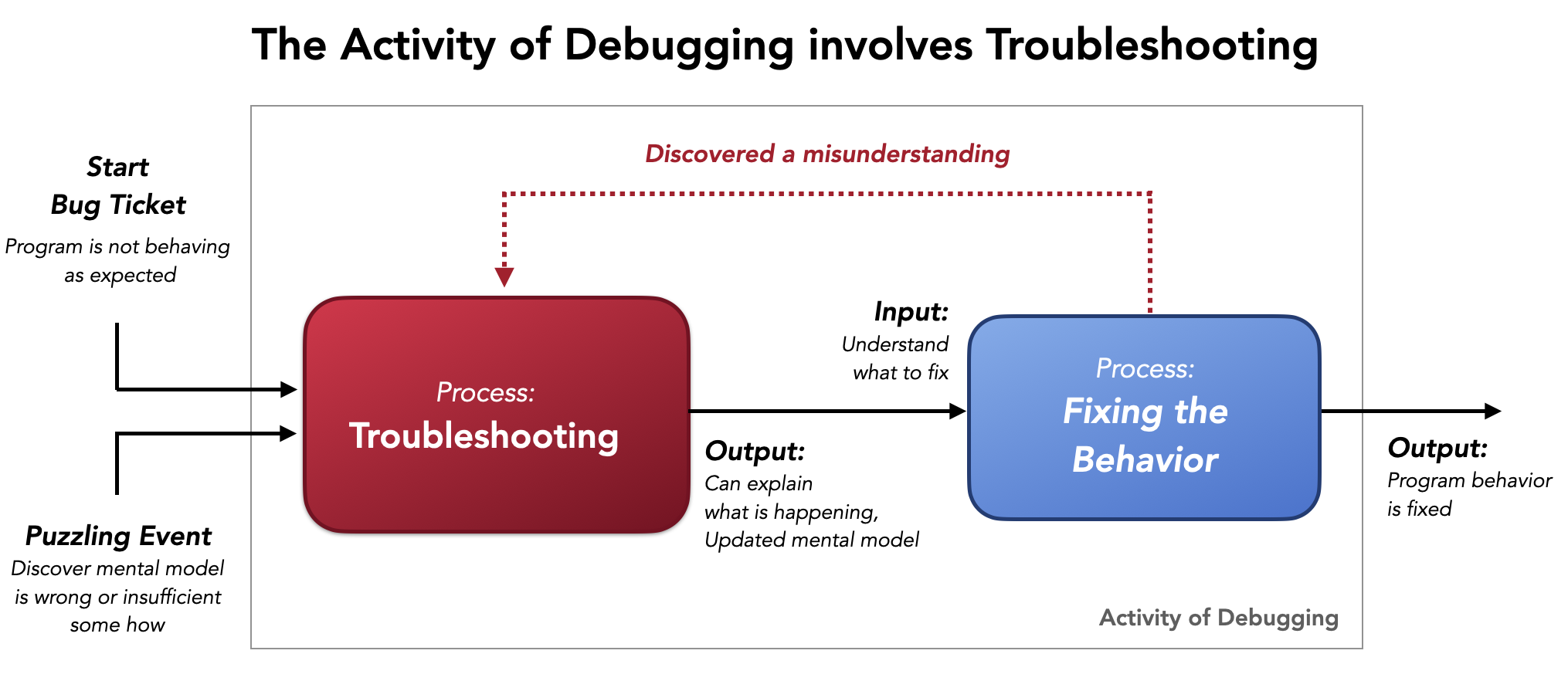}}
  \caption{The activity of debugging is a composite process that includes Troubleshooting and Fixing the Behavior. Developers often loop between these phases as their understanding evolves.}
  \label{fig:trouble_debug}
\end{figure*}

What makes \rev{the developer's} process \textit{troubleshooting} is not the presence of a bug, {but rather} the developer’s engagement in a cognitive process of trying to understand unexpected behavior.  \rev{These are related, but oriented differently: one reflects a condition of the code, the other a shift in the developer’s cognitive activity.} To help clarify how \rev{troubleshooting} fits within the broader activity of debugging, Figure~\ref{fig:trouble_debug} shows how we conceptualize the activity of debugging as a composite process that includes both troubleshooting and fixing the behavior. When developers either start a bug ticket or encounter puzzling behavior, they begin troubleshooting.  Once they believe they have understood the issue and can explain the behavior, they move into implementing a fix. If the fix does not work or reveals a misunderstanding, they return to troubleshooting. Developers can cycle through this loop multiple times as their understanding evolves.

In the sections that follow, we share how the developer’s cognitive experience unfolds\rev{—not as discrete sequential stages, but through a set of theoretical categories that reflect the layered and interwoven aspects of lived experience.}

\subsection{The Confusion Experience} \label{confusion_experience}

Developers described the initiation of troubleshooting as a shift in cognitive mode, often triggered by encountering unexpected system behavior.  Across interviews, developers used a range of terms—``confusing,'' ``puzzling,'' ``surprising,'' and occasionally ``fuzzy''—to describe their experience. We group these descriptions under the term \textit{confusion experience}, which begins when the developer suspects that something is not behaving as expected and their current mental model is insufficient to explain it. \rev{This marks the first major transition in our model.}

\change{In this section, we explore the cognitive experience of confusion through six interrelated patterns. We begin with \textit{what confusion feels like}, where developers describe the encounter with unexpected behavior. We then discuss how \textit{confusion stirs in the background} even when the developer is doing other activities and insights can emerge. Next, in \textit{confusion is fatiguing}, we describe the mental strain that builds when confusion remains unresolved.  Then, we discuss \textit{confusion as a cognitive context switch} and the nature of the attention shift caused by confusion.  Finally, we discuss nuances and variations of the confusion experience and the way \textit{familiar patterns feel less confusing}.}

\subsubsection{What Confusion Feels Like}

\change{At the initiation of troubleshooting, developers described a similar phenomenon using a variety of words.  P10 reflected on the discomfort of realizing their mental model was wrong:}
\begin{quote}
\textit{``It was basically that nagging feeling. I realized I felt that my mental model was the thing that was wrong somehow, but the reason it was wrong was elusive because I was fairly confident that my mental model was true. So that was just a very frustrating feeling.''}
\end{quote}

\begin{figure*}[h]
  \centering
\includegraphics[width=\linewidth]{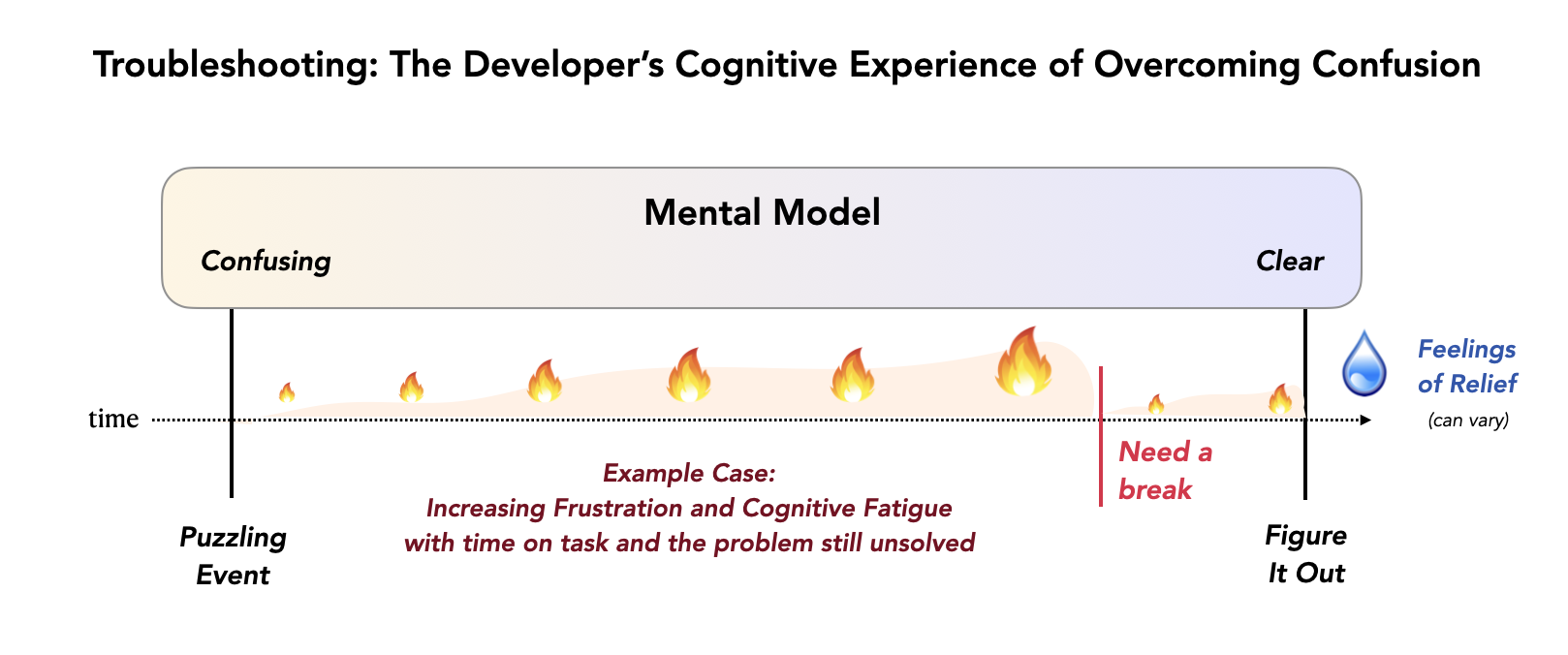}
  \caption{Model of the developer's cognitive experience during troubleshooting, highlighting the dynamics of increasing frustration and cognitive fatigue over time, followed by relief upon figuring it out, and the mental model going from confusing to clear.}
\label{fig:confusion_experience}
\end{figure*}

\change{This moment of discovering that one's mental model is insufficient but not knowing why marks the beginning of the \textit{confusion experience}. Developers encountered this experience in a range of scenarios, most often when observing unexpected output during program execution. P14 shared a story of feeling confused after encountering unexpected data in the database when running a data migration job:}
\begin{quote}
\textit{``I saw that some of the [rows] that the original job should not have been able to write to, did have the backfilled data. And that was very confusing for me. I was like, no, no, no, my whole model of why this category, this class of data, hasn't been written, hinges upon the fact that I have this bit set on them. And I know it's set on these, but why do some of them actually have the data backfilled? So that's what set me off on the problem. And so it was like, okay, something doesn't behave the way I thought it did.  How am I gonna figure this out?''}
\end{quote}
\change{These accounts illustrate how the confusion experience is triggered by a mismatch between the developer’s expectations and the system’s behavior—a recognition that something is off, but without a clear explanation. The developer’s attention shifts toward understanding the cause of the behavior and reconstructing a coherent mental model of what is happening.} \change{Figure~\ref{fig:confusion_experience} presents a model of the developer’s cognitive experience of overcoming confusion during troubleshooting. The \textit{puzzling event} initiates troubleshooting as the developer begins to wonder why the behavior is happening. The model depicts the flow of experience from this initial puzzling event, through a period of confusion, to eventual resolution and greater clarity. It also summarizes the emotional arc developers reported during the process, which we continue to explore in later sections. }

\change{The model highlights a common pattern of increasing frustration and cognitive fatigue with increasing time on task—an experience that resonated strongly with participants.  The developers often described taking breaks as a useful strategy to relieve frustration, with some noting that problems were easier to solve after stepping away. The specific dynamics of cognitive fatigue we discuss in more detail in Section~\ref{cfatigue}. } 

\change{Cognitive science offers several models that help explain the confusion experience as a response to a violated expectation \cite{pinquart_why_2021}.}  For example, the Predictive Processing (PP) framework assumes the brain's overarching goal is to minimize surprise (expectation violations and prediction errors), and when encountering a violated prediction, the brain will invoke cognitive processes to reduce future prediction errors \cite{clark_whatever_2013}.  Similarly, the Meaning Maintenance Model (MMM) \change{proposes that humans rely on coherent mental representations to organize perception, and violations of those representations lead to a state of aversive arousal and compensating efforts to restore meaning and reduce arousal \cite{heine_meaning_2006}.} 

\change{Kahneman described how a violated expectation creates an  \textit{orientation response}, an involuntary reaction to a novel stimulus that maps to physiological changes and shifts in attention \cite{kahneman_attention_1973, sokolov1963perception}.}  This shift includes four key components: 1) enhancing processing of the novel stimulus with additional attention, 2) inhibition of any irrelevant activity, 3) orientation toward probable sources of new information, and 4) elevated arousal to support the new attentional demands \cite{kahneman_attention_1973}. \change{The onset of confusion also resembles what Duraes et al.\ observed in neuroimaging studies: increased activation of the insula during the moment the developer first suspects a bug, a brain region associated with uncertainty and attentional control  \cite{duraes_wap_2016}. While our \rev{models were} developed independently, this alignment suggests that the confusion experience may have identifiable neural correlates, marking a cognitive shift characterized by uncertainty, attentional reorientation, and heightened executive processing.}

\subsubsection{Confusion Stirs in the Background}

\change{Another key characteristic of the confusion experience is that it continues to stir in the background, even as the developer turns attention to other tasks. Participants described insights emerging spontaneously during unrelated activities, suggesting that cognitive resources may continue working to resolve the puzzle beneath the surface of conscious awareness.  These moments offer clues into the type of insights generated by this background process: often partial, imprecise, or associative ideas that point toward something worth exploring.} For example, P15 described an insight that arose while walking to lunch:
\begin{quote}
\textit{``So this just popped into my head when I was walking down for lunch earlier today. It's like, oh, I've got to remember that. I'm going to look into that as well.''}  
\end{quote}
\change{The insight was not a fully formed hypothesis, but an awareness of a constraint that might be affecting the behavior—something to investigate. Similarly, P10 described an insight that emerged while taking a shower the morning after a frustrating troubleshooting session:}
\begin{quote}
\textit{``I was really banging on it, and banging on it, wasn't getting anywhere. This was Friday evening, so then I went home, went to bed.  Next morning, in the shower after the gym, I had an insight that there's two levels of validation that we could do for the thing that we were doing, and the lower-level validation, I was never calling that, because generally that wasn't giving useful results. It was just this other level validation, was the one I was reporting. So the existence of that other validation kicked into my head. So it's like, oh, what happens if I look at that validation? So then, I don't normally work on the weekends, but because I had this insight, I was like, I need to test to see if that's what it was.''}
\end{quote}
\change{In both cases, the background stirring of confusion gave rise to insights that were actionable but indirect: not definitive answers, but new investigative paths that might reveal the cause of the issue.} Interestingly, P10 could not resist testing the strategy immediately, even though it was the weekend. \change{Their story highlights how unresolved confusion can create a lingering sense of tension—a motivational pull to resolve the uncertainty and restore coherence. If the strategy worked, the background stirring of confusion would subside, and the developer’s attention could fully return to other tasks. This example raises important questions about the cognitive cost of unresolved confusion, and how its presence in the background might impact attention, motivation, and the ability to focus on other work.}

\subsubsection{Confusion is Fatiguing} \label{cfatigue}

\begin{figure}[h]
\centerline{\includegraphics[scale=0.53]{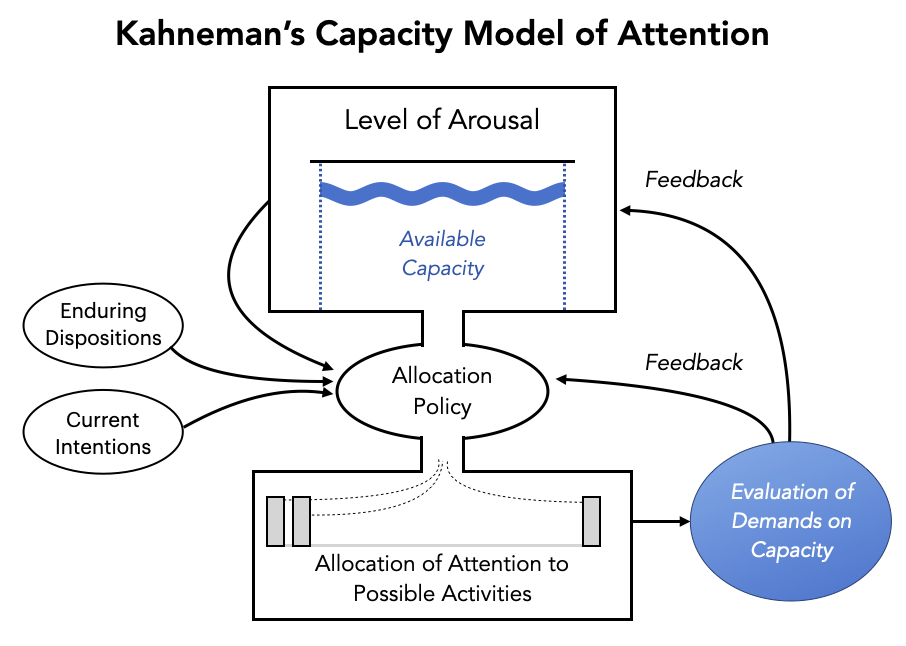}}
\caption{Redrawn capacity model of attention by Daniel Kahneman  \cite{kahneman_attention_1973}}
\label{fig:kahneman_attention}
\end{figure}

Another key aspect of confusion is the experience of feeling cognitively tired\change{, beyond typical feature work.}  We asked P25 how they felt after troubleshooting a problem:
\begin{quote}
\textit{``Usually relieved, ready for a break. I think being in that headspace is quite tiring. And so often when there is a breakthrough, like in a problem being solved, and sort of a crisis is averted, it's nice to just take a break.''}
\end{quote}
P25 reported that the \textit{headspace} of troubleshooting was tiring\change{, suggesting that it is the sustained mental absorption in a confusing problem context that contributes to fatigue. Other participants echoed this pattern. P04 described that additional attentional effort was needed to troubleshoot effectively, and P07, P15, P19, P20, and P24 all emphasized the need to take breaks—suggesting that the confusion experience carries a cognitive cost that accumulates over time. P25 also described how the effects of confusion accumulate over several days:}
\begin{quote}
\textit{``I was a little frustrated. I think 'cause I was tired, 'cause I'd been working on this all week. It was like 4 p.m. on a Thursday that we were doing this. And I just, I did not have any patience left for it to not work. (laughs) So that definitely put my blinders on, and that's where having my coworker come in calmly, not at all sick of working on this the way I was, was really helpful.''}
\end{quote}
P25's description indicates the emotional urge to stop as described by Ackerman \cite{ackerman_cognitive_2011} indicating cognitive fatigue.  \change{Although an individual might feel an urge to stop, they can still continue working by exerting compensatory effort--but at the cost of increased physiological stress \cite{ackerman_cognitive_2011}.}  \change{P07, P10, and P13 also spoke of blindness effects while coding.} P25 elaborated on what the experience is like:
\begin{quote}
\textit{``Just literally being tired, and being certain that I am reading a chunk of code that says one thing, and it does not do what I think. I'm just kind of skimming it. Much like those exercises, you see those shared memes where if you remove all the vowels, you can still understand the words. [...] It's just literally not seeing an issue.''}
\end{quote}
\change{This \textit{blindness effect} resembles the perceptual breakdowns associated with high cognitive control tasks and cognitive fatigue\cite{boksem_effects_2005}.}  \change{Kahneman’s capacity model of attention helps explain this fatigue effect mechanistically  \cite{ackerman_cognitive_2011} \cite{kahneman_attention_1973}. The model views attention as a limited resource that fluctuates with arousal and task demands (Figure \ref{fig:kahneman_attention}) \cite{kahneman_attention_1973}.   If, as our findings suggest, confusion initiates an orientation response, this would increase arousal and draw more attention to the unexpected stimulus—thereby accelerating the depletion of attentional resources. When attentional capacity is exhausted, the individual can no longer sustain the focus that the task requires, leading to fatigue and a need to rest \cite{ackerman_cognitive_2011}.}

\change{While \rev{taking} breaks are often framed as a response to fatigue, developers also described them as an active strategy for becoming more effective at troubleshooting. P15 described the power of stepping away from the problem:}
\begin{quote}
\textit{``Sometimes you get so locked in to what you're doing. And I mean, this is a real--this is like a wizard move, is to just step away. Go out and get some coffee. It totally helps. I've found that many times over in my career.''}
\end{quote}
\rev{Several developers found} that taking breaks helped them get unstuck. The high attentional demands of troubleshooting help explain both the rapid onset of fatigue and the emphasis on taking breaks as a practical strategy.

\subsubsection{Confusion as a Cognitive Context Switch} \label{context-switch}

\begin{figure*}[h]
  \centering
\includegraphics[scale=0.53]{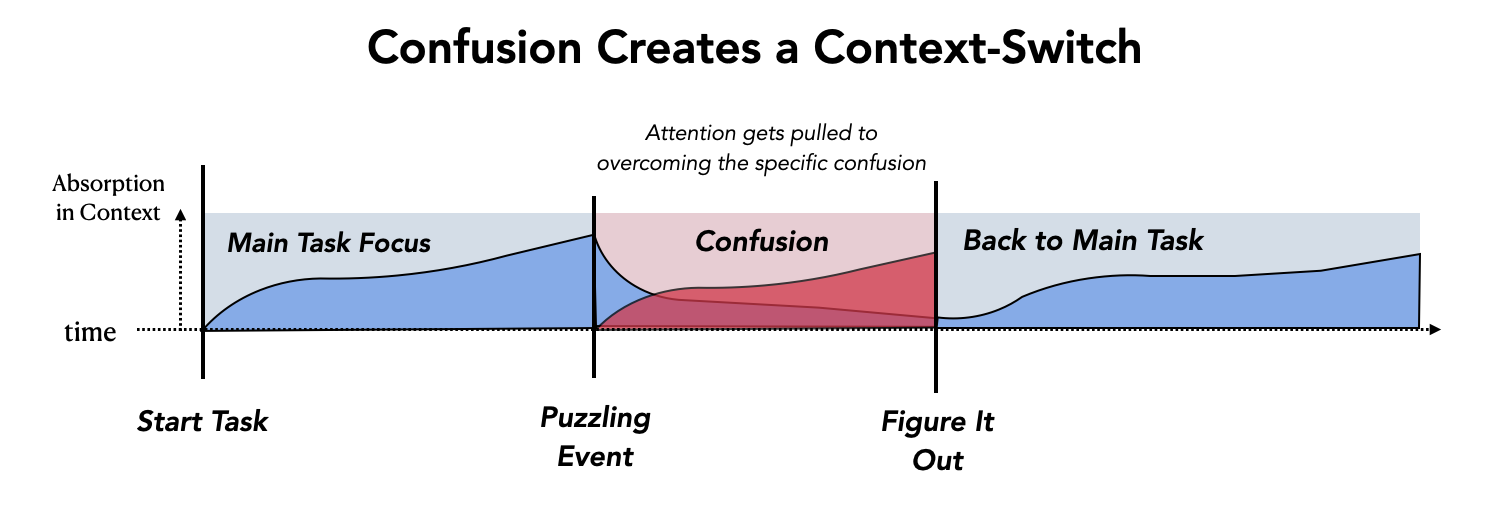}
  \caption{Model of how the orientation response shifts the developer's focus by pulling attention to the challenge of overcoming the confusion, creating a cognitive context switch.  If the troubleshooting takes a while, once the developer resolves the confusion and resumes their main task, it can take time and effort to recall the previous context.}
  \label{fig:context_switch}
\end{figure*}

A key clarification about the confusion experience emerged when we began modeling the specific moment when troubleshooting begins. While P23 initially described confusion as something that ``breaks your flow,'' follow-up interviews revealed a more nuanced picture. P10 and P14 clarified that their flow was not actually ``broken''—they still felt immersed in a flow state—but the experience was more like a type of context switch.  \rev{P23 distinguished between two different situations of encountering unexpected behavior to highlight when troubleshooting actually begins:}
\begin{quote}
\textit{``In the first one, it feels like they just kind of fit right in with the regular rhythm.  It's like wait, that's wrong. I look at the data, I know why it's wrong, move on. Sometimes I look at the data, I don't know why it's wrong, and now, I need to stop and start reading back, re-evaluating what I know.''}
\end{quote}
P23’s story clarifies an important point: confusion is not simply triggered by the presence of a bug, but by a felt sense of puzzlement. If a developer sees an unexpected behavior but immediately understands its cause, the fix flows naturally within the main task rhythm. However, if the behavior does not make sense—if the expected fix fails or the cause is unclear—the developer experiences confusion. It is this ``wait, what?'' moment that initiates the shift into the confusion experience and launches the cognitive process of troubleshooting.  \rev{Figure~\ref{fig:context_switch} presents the final model of the cognitive context switch, co-developed by participants and researcher, that describes the experience.}

\change{During the follow-up interviews, P15 described how the original task context is not instantly lost the moment confusion begins, but fades over time as the developer becomes increasingly absorbed in the troubleshooting context. After resolving the confusion, developers do not start from zero, but they do need time to recall and rebuild the context to return to the original task. This re-immersion process echoes the ``resumption lag'' effect found in research on interruptions and context switching \cite{borst_what_2015}.}

\subsubsection{Familiar Patterns Feel Less Confusing} \label{familiar-patterns}

\change{A subtle but important nuance emerged in follow-up interviews with P10 and P15: unexpected behavior does not always evoke a felt sense of confusion, even when it initiates troubleshooting. For example, if an error message is familiar—even if the developer does not know the exact root cause—it may still be clear where to look and what to look for. In these cases, troubleshooting proceeds without the disorientation associated with confusion. It is when initial assumptions fail and the cause remains unclear that the experience feels truly confusing.}

\change{This distinction highlights a spectrum of troubleshooting experiences. When errors follow familiar patterns, and the developer has a mental template for resolving them, the process feels more routine—even if it still involves troubleshooting. By contrast, novel or unexpected failures—those that do not conform to familiar patterns—are more likely to produce the experience of confusion.  P27 added further nuance, noting that when the goal is to fix a bug from a ticket, rather than responding to an unexpected issue during creation, the experience feels different.  It shares elements of puzzlement, but the orientation toward the task is different, and it was not clear what the right word for that experience would be.}

\change{While these distinctions complicate attempts to neatly model the confusion experience, they are essential for understanding its phenomenology. As a constructivist grounded theory, this work does not aim to impose clean categorical boundaries, but to articulate patterns that resonate with lived experience. The blurred edges—between puzzlement and familiarity, between confusion and fuzziness—point to valuable directions for future research.}

\subsection{Trying to Gain Clarity} \label{gain_clarity}

Troubleshooting can be understood as a process of striving, of \textit{trying to gain clarity}\rev{—a cognitive process initiated in order to overcome confusion.} \rev{In this section, we examine the thought processes of the developer and the iterative strategy for gaining clarity and making progress.} We begin with \textit{clarity is the goal}, where developers articulate the felt aim of troubleshooting as achieving clarity. We then describe \textit{thinking of things to try}, \rev{a generative loop of identifying possible actions that might illuminate the problem.} Next, in \textit{the notion of progress}, \rev{we examine how developers recognize movement toward clarity, even in the absence of a solution.} Finally, in \textit{talking it out}, we highlight how elucidating the problem can surface new insights.

\subsubsection{Clarity is the Goal}

\change{Across participants, a common thread emerged: a goal of trying to achieve clarity.  Rather than an immediate leap from confusion to clarity, developers described an iterative effort toward understanding. P19 described how troubleshooting gradually improved the clarity of their thinking, beginning with uncertainty, and ending with a clear understanding. While this \rev{phenomenon} was implicit across many stories, P06 articulated it most explicitly:}
\begin{quote}
\textit{``I think the main focus of what I'm trying to do is achieve clarity. What is the problem? Do I understand the problem? Do I have an idea of which parts are affected?''}
\end{quote}
\change{Other developers echoed this same striving for clarity in diverse ways. P03 described stepping back to consider the meta problem and using stream-of-consciousness notes to track emerging understanding.  P22 and P23 used detective-style notes to synthesize evidence. P06 described a methodical process of gaining clarity piece by piece. P01 and P07 emphasized getting clearer before asking for help. Across these accounts, clarity stood out as a central goal.}

\subsubsection{Thinking of Things to Try}

\begin{figure*}[h]
\centerline{\includegraphics[scale=0.5] {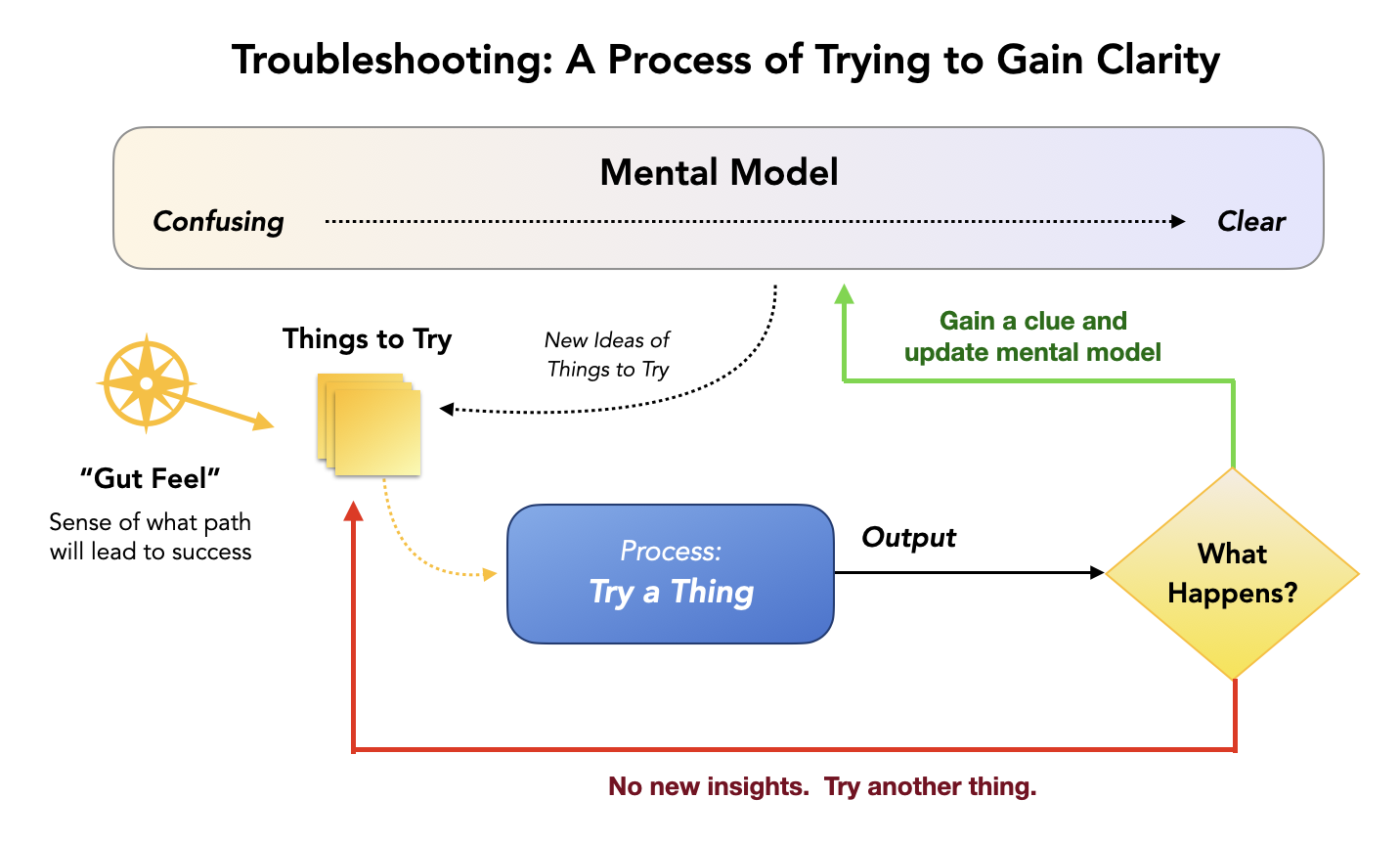}}
  \caption{Model of the cognitive process of troubleshooting, a striving of trying to gain clarity, to overcome the confusion.  The model highlights the iterative process of making progress by trying things in an effort to gain clues. }
  \label{fig:try_a_thing}
\end{figure*}

In the process of trying to gain clarity, the developer begins \textit{thinking of things to try}. A ``thing to try'' is an imprecise pointer to an imprecise thought—an actionable next step that might help the developer make progress. \change{Figure~\ref{fig:try_a_thing} describes the iterative process of trying to gain clarity through a trial and experimentation process modeled in the language of developers.} The figure represents these tentative ideas as a pile of sticky notes: possible actions, imagined hypotheses, or strategies for gathering more information.

\change{P15 described their thought process during troubleshooting and their evolving mental list of things to try:}
\begin{quote}
\textit{``What else is going through my head? Just all the possibilities—the things I could do. I'm thinking, well, maybe I should contact this group. No, I shouldn't because there's some things I want to try first. You're constantly trying to form a plan that makes sense. And usually that plan is small enough that it can fit in your head.''}
\end{quote}

\change{Developers described these options as ideas, steps, or investigative paths. P27 emphasized how having steps to try helped them stay engaged in the process:
\begin{quote}
\textit{``I was engaged in the step-by-step process. So for me, the actual having steps to try, and then checking if it worked, and then going back and checking the next step or trying this and then trying another step and seeing if that worked. I guess that was my curiosity. I wonder if this will work.''}
\end{quote}}

\change{The experience of feeling stuck was closely tied to running out of things to try. Several developers preferred to wait until they had run out of things to try before asking for help.} \change{P15 and P10 shared a strong preference of only asking when feeling ``really truly stuck.''  As long as developers had things to try, they did not necessarily feel stuck—even if  still confused.  In this way, the presence or absence of things to try shaped the developer’s perception of stuckness.}

\subsubsection{The Notion of Progress}

\change{During troubleshooting, }the feeling of making progress comes from gaining additional knowledge—knowledge that helps the developer gain clarity.  \change{P10 described the feeling of progress:}
\begin{quote}
\textit{``[There is a] feeling like you're actually making progress through it. So if I'm troubleshooting and trying things, and if it's like dead end after dead end, and you're always going back to square one—to me that's very difficult, because you're not gaining any additional knowledge in that experience. So you're not actually ruling anything out. Like if you go down a dead end, and then that dead end, you actually learned that, oh, this lets me exclude a set of the potential solution space of what it could be, then you did actually learn something.''} 
\end{quote}
\change{Progress, then, is not only about finding the solution, but also about ruling out possibilities and shrinking the possible solution space. P10 further distinguished between two types of knowledge that can move troubleshooting forward:}
\begin{quote}
\textit{``I always want to be somehow gaining some bit of knowledge with each step that I take, whether it is positive knowledge that is telling me a fact about it, that might be honing me in, or negative knowledge that is exclusionary and taking away. It's still reducing the solution space that I'm dealing with.''}
\end{quote}
\change{Figure \ref{fig:try_a_thing} shows this feedback loop of making progress when gaining a clue (either positive or negative knowledge).  When no new information is discovered, however, the developer does not make progress, and must try another approach.
The notion of progress centers around gaining new information that brings greater clarity.}

\subsubsection{Talking It Out} 

There is a common practice during troubleshooting that developers call ``rubber ducking,'' where a developer explains the problem to a rubber duck (or a teddy bear) as a way to gain clarity.  \change{P03, P04, P13, P14, P20, and P23 all highlighted the effectiveness of rubber ducking.} P25 described the effect:
\begin{quote}
\textit{``So I'm a really huge fan of what most people call the duck effect, but I've always called it the teddy bear effect, because I had a prof at university who once said, `you know, you can come ask me questions about your assignment, but you have to explain it to the bear on the filing cabinet first. And if the bear can't solve your problem, then I'll talk to you.' And it was just an illustration of that action of, if you describe your issue out loud, often it will occur to you.''} 
\end{quote}
\change{Developers described many variations of this practice. P23 wrote draft emails to hypothetical developers. P04, P14, and P15 preferred talking to real people, but still found that articulating the problem often brought the insight they needed. P25 offered a key insight on why the technique was so effective:}
\begin{quote}
\textit{``I think what it is, is it's not so much that I'm asking questions, it's that I'm preparing to explain it to someone else as if they have no idea what I'm talking about. And it's that act of organizing your thoughts so that you can explain them to someone else, so that they make sense to an external person.''}
\end{quote}
In trying to make the problem clear to someone else, a developer must first make it clear to themselves. In this way, \textit{talking it out} \rev{is} not just a communication tool, but a powerful cognitive tool that can lead to breakthrough insights.

\subsection{Poking and Seeing} \label{poking_and_seeing}

\rev{While the previous section focused on the developer's thought process, this section centers the developer’s hands-on engagement with the system. \textit{Poking and seeing} is a process of hands-on experimentation in service of gaining clarity, where developers take targeted actions and observe system behavior to generate clues and insight. This process is shaped not only by what the developer does, but also how the system responds—what it reveals, how easily it can be manipulated, and what signals are available to guide interpretation.}

\rev{We examine three interwoven aspects of this interactive process.} We begin with \textit{poking at the problem}, where developers describe how they craft feedback loops to trigger informative behavior. We then turn to \textit{seeing what is happening}, which highlights how developers refine their understanding through observation. Finally, in \textit{frustration vs confidence}, \rev{we explore how the presence or absence of supportive infrastructure shapes the developer’s emotional experience and sense of agency.}

\subsubsection{Poking at the Problem}

\change{\textit{Poking at the problem} is a strategy developers use to generate clues by crafting feedback loops that exercise specific parts of the system. The goal is to trigger behavior that sheds light on the issue at hand.} \change{P10, who works in a large organization with sophisticated automated tooling, described this approach in detail:}
\begin{quote}
\textit{``I try to craft, if possible, a tighter feedback loop for the problem in order to poke at it. So this is sometimes a debugger, but not often a debugger. So if it is a code-based problem that I have, I will often go back to tests.'' }
\end{quote}
\change{In this account, a ``poker'' could be an automated test, a debugger, or any targeted mechanism for running a subset of the system and inspecting the output. Automated tests were mentioned as essential tools for supporting troubleshooting, enabling developers to isolate the behavior of interest.} P10 further described a form of ad hoc tool construction to create a more effective feedback loop:
\begin{quote}
\textit{``Let me write a targeted test that is only focusing on that bit of code, and see if I can get that to fail in the same way, and then start to kind of attack that problem. So I would maybe generalize that as ad hoc tool construction based upon the things that I have, in order to very quickly poke at the failure that I'm seeing, and enable me to iterate more quickly towards a solution.'' }  
\end{quote}
\change{By constructing ad hoc targeted tests that isolate a smaller unit of code, the clues can be easier to interpret and help the developer gain clarity more quickly.} 

\subsubsection{Seeing What is Happening}

Once the developer has a feedback loop \change{in place, the resulting output supports the developer in \textit{seeing what is happening}. By making small, deliberate changes and observing the effects, developers can run experiments to uncover how the software behaves—and why it might be failing.} P04, \change{who works as a technical coach and consultant,} described the approach:
\begin{quote}
\textit{``I think what underlies a lot of this is very much an experimental approach. Let me change one tiny thing and see if that changes things. It did or it didn't. It didn't. Okay, then this is not part of it. Let me change one other tiny thing.  Because what I see a lot of, is let me change a whole bunch of things, all at once.''}
\end{quote}
\change{An important aspect of designing effective experiments is ensuring that the clues are unambiguous. If the behavior changes after a single modification, the developer can confidently attribute the effect to that change. But if multiple changes are made at once, the signal becomes noisy, and the developer risks drawing the wrong conclusion and setting themselves up for future mistakes.}

\rev{Deployed environments—especially production—can make it harder to see what is happening.} P02, who works in a large organization on the developer experience team, highlighted the difficulty of seeing in production:
\begin{quote}
\textit{``Once you start running something in production, it gets really hard to tell what's happening, even if you have an observable platform, at least without a lot of time and energy and money. It becomes really hard to tell what's actually going on there.''}
\end{quote} 
\change{Observability infrastructure—logs, dashboards, and telemetry—can make it easier to see, but these systems can require significant effort to set up and maintain.} \change{P14, who works in a large organization on developer experience tools, emphasized the importance of instrumenting code so that it tells a clear story about its behavior:}
\begin{quote}
\textit{``When you write software, instrument it, so that when you're looking at the results of that software having run, you can see a story in the data. You can see telemetry information about requests per second, and all that kind of stuff in monitoring dashboards, and logs. Leave yourself clues about what the software is doing, instead of it just being a black box, and you don't know what's going on inside. [...] It would be great to just sort of shout from the rooftops, 'Write better error messages!' And I think that  would help a lot.''}
\end{quote}
\change{P14 offered three strategies for making \rev{deployed} systems more legible: 1) leave clues in the code and output, 2) tell a coherent story with telemetry, logs, and dashboards, and 3) write better error messages.} \change{P16, who works in a small organization in a technical leadership role, highlighted a different strategy—investing in tools that help developers reproduce production issues locally:}
\begin{quote}
\textit{``I load up my workspace, and I can point my project to the same environment that it was being reported on, so I have the right data.  And basically, the steps involved there are, I run a script which points my environment to that same one that [the bug] was reported on, then I can initiate a targeted action, like an execution action in my IDE to basically debug locally against that environment.''}
\end{quote}
P16 was able to troubleshoot more quickly because of an investment in automated tooling; tooling that helped \rev{them} reproduce the problem locally.  Despite the time investment in building such tools, P23 highlighted the importance:
\begin{quote}
\textit{``I'd say one of the biggest things is probably supporting developers in being able to see what their system is doing. Troubleshooting is often simple, if you can just see what's happening.'' }
\end{quote}
\rev{The developer's} ability to see what the system is doing emerged as a central enabler of effective troubleshooting.

\subsubsection{Impact of Tooling: Frustration vs Confidence} \label{frustration}

\change{The emotional experience of troubleshooting can vary dramatically} depending on the presence or absence of automated tooling and infrastructure that makes poking and seeing easier. By contrasting developer experiences between healthy environments versus challenging environments, we can get a sense of what makes developers feel confident versus frustrated during troubleshooting. 
P11, who works in a midsize organization as a front-end developer, described a frustrating legacy code environment where troubleshooting is often difficult:
\begin{quote}
\textit{``No one knows, nobody knows, literally, like nobody knows why, how does that work? But you're not given the time to understand any of it. So because so much of it was out of my control, and so much of it just had to do with the fact that we're working with this legacy code base, this would be an eight [difficulty] because you feel very helpless. [...] The problem is we are not given time. [...] Anything that will require an investment of time won't be implemented in most companies because you're trying to maximize profit.  So making it easier for the developer to develop is not a priority at most companies, even if they will verbally say that it is.''}
\end{quote}
\change{In a follow-up interview, P18 connected these feelings to a perceived loss of control}. While investing in automated tools and infrastructure does not add features to the product, the investment in making poking and seeing easier can drastically reduce the difficulty and frustration (and cost) of updating the software. By contrast, P16 described a confident developer experience that feels good while troubleshooting:
\begin{quote}
\textit{``We worked really hard to make it have a good developer experience. And all those tools that I described, make it relatively fast and easy to point my environment from A to B, and to then be able to launch the debugger, go through steps like getting security authorization tokens, and all that kind of stuff. [...] And the reason it makes me feel good is because I've worked long enough in other organizations to know, that something like that could take a whole morning.''}
\end{quote}
Automated tooling can have a significant impact on the Developer's Experience, and turn a stressful frustrating experience into one that feels good.  \change{These contrasts help illustrate how tools and infrastructure shape not only the developer’s effectiveness, but also their emotional experience—reinforcing the role of infrastructure in enabling intuitive and confident troubleshooting.}

\subsection{Experiential Intuition} \label{experiential_intuition}

\rev{As developers strive to make sense of a confusing system behavior, they also draw on a ``gut feel'' intuition that guides their strategy, shaping where they look and how they begin. This intuitive sense—what we call \textit{experiential intuition}—is a tacit form of knowing shaped by accumulated experience (expertise), in which perceived similarities to past situations provide a felt sense of direction, even before a clear rationale has formed.}

In this section, we introduce the concept of experiential intuition and qualitative accounts of how this intuition is experienced.  We begin with \textit{experiential intuition guides the strategy}, where developers describe how pattern-matching against past experiences provides a felt direction for troubleshooting. In \textit{jumping to the wrong conclusion}, we examine how these intuitions—while often helpful—can mislead, especially when a problem appears familiar but differs in crucial ways. Finally, in \textit{the importance of familiarity}, we highlight how familiarity shapes the effectiveness of experiential intuition, and how a developer’s past experiences influence their troubleshooting ability.

\subsubsection{Experiential Intuition Guides the Strategy}

\begin{figure*}[h]
\centerline{\includegraphics[scale=0.42]{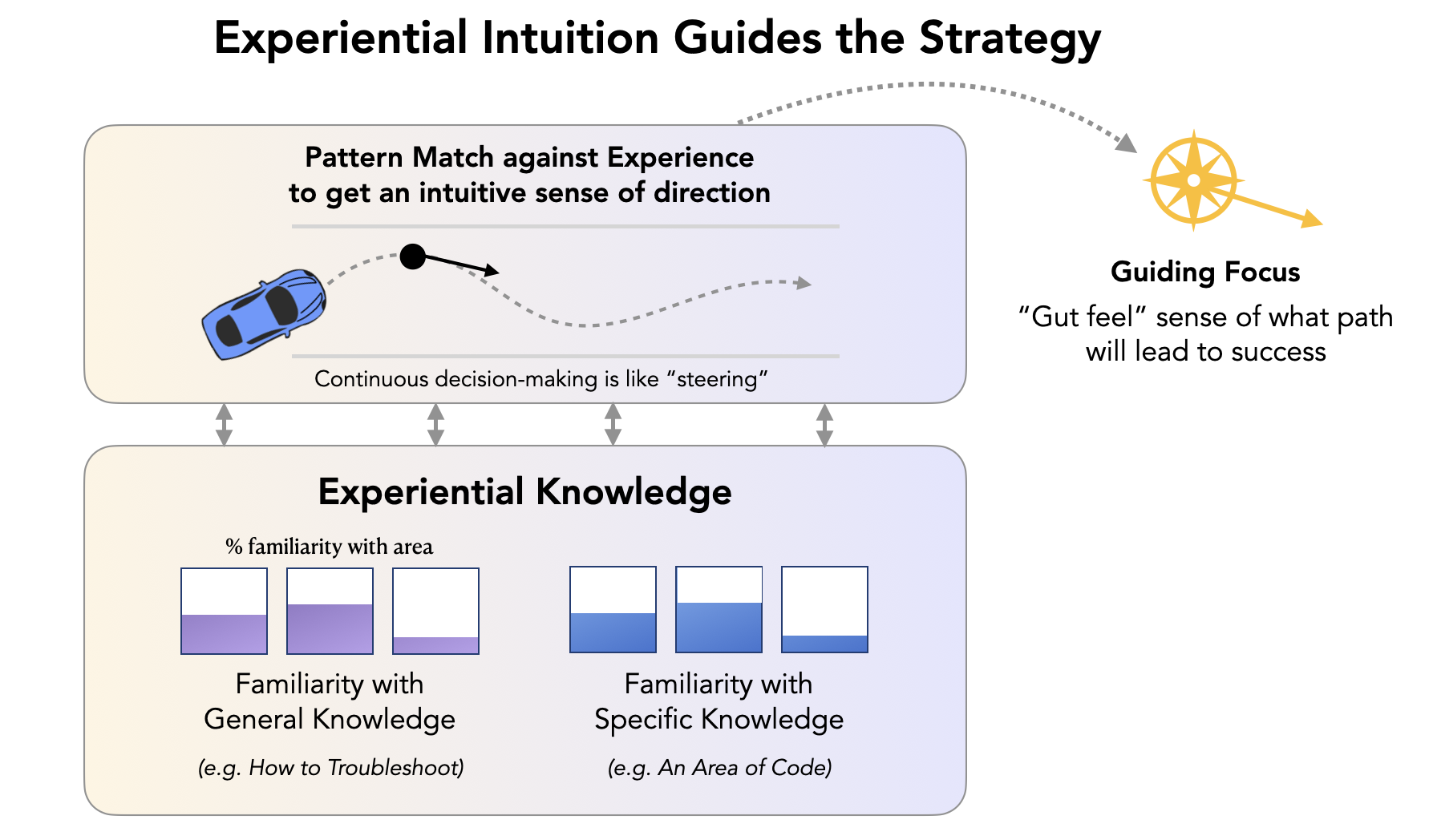}}
  \caption{\rev{Experiential intuition guides troubleshooting strategy by drawing on the developer’s accumulated experiences to create a ``gut feel'' for what path might lead to success. This intuition emerges from continuous pattern-matching against experiential knowledge—both general (e.g. how to troubleshoot) and specific (e.g. familiarity with a particular codebase). As familiarity increases, the developer’s intuitive steering becomes more confident and effective.}}
  \label{fig:experiential_intuition}
\end{figure*}

We asked P10 what sort of tools and strategies they normally use, if any, to enhance their ability to troubleshoot, and learned about the phenomenon of experiential intuition:
\begin{quote}
\textit{``There's a certain amount of intuition... that is the first word that came to mind, but I think that's not the best phrase. It's probably more something more experience knowledge, experiential-based, based upon what I've seen before, like this problem smells like this, sort of thing. And that kind of guides the strategy.''\footnote{Inspired by the notion of code smells \cite{fowler2018refactoring}, indicating subtle cues that something may be wrong even without a clear error.}}
\end{quote}
Experiential intuition creates a felt sense of \change{where the problem might be and how to proceed.} Figure \ref{fig:experiential_intuition} summarizes how experiential intuition emerges from experiential knowledge and pattern-matching the current situation with past experience to result in a ``gut feel'' sense. \change{We relate this sense of direction to a driving metaphor, where intuition functions like steering with moment-to-moment adjustments in direction.} P25 describes how experiential intuition guides the strategy:
\begin{quote}
\textit{``And I was like, 'Ah, what's happened?' And I had guesses. This is something where I maybe went wrong.  When I was about to do the deploy, I was concerned about the interaction with the database because I had so much trouble setting it up. And there were so many aspects of it that were new. And I just was really uncertain about it. So I kind of went there first, instead of going to the logs, which my coworker did, and I immediately saw the problem.''}
\end{quote}
Experiential intuition helped the developer \rev{make a good guess about where the problem might be, thereby enabling faster progress.}

\subsubsection{Jumping to the Wrong Conclusion}

\change{One side effect of relying on experiential intuition is that intuitive hunches can sometimes lead developers astray.} P04 made the connection between experiential intuition and accidentally solving the wrong problem:
\begin{quote}
\textit{``It happens to me quite a bit of solving a problem that's not the one that you have, because it looks a lot like ones you've seen before. And so, because we tend to pattern match against our experience, it's like, oh, this is like this other thing. And then it's not at all like this other thing, because you missed something.''}
\end{quote}
\change{Because experiential intuition is shaped by pattern-matching against past experience (Figure \ref{fig:experiential_intuition}), it can lead to incorrect assumptions when a new problem appears similar to a previous one but differs in important ways. One of the ways developers counter this risk is a discipline of questioning assumptions.  P15 shared this wisdom:}
\begin{quote}
\textit{``What you have to do is question every assumption that you have about how things work. That is key in finding out what the solution to a problem is. Because usually, it's something that you're assuming is working a particular way, that is not working that way. And you'll investigate all the way around it without zooming into that and saying, let me just verify that this is truly the case.''}
\end{quote}
P04 and P25 also highlighted the importance of questioning assumptions as a core part of their troubleshooting strategy.

\subsubsection{How Experiential Intuition Develops}

\rev{Developers must learn how to troubleshoot software systems, yet troubleshooting is not a skill that is typically taught directly. P10 highlights the indirect way troubleshooting is learned:}
\begin{quote}
\textit{``You learn how to code, or learn how to solve problems, but you don't necessarily learn troubleshooting.  There's no explicit teaching around that.  It's all very, very experiential. [...] If you had a more generalist background in doing more things, touching different types of systems, different layers of software stacks and building up a better mental model of how everything works, it could then guide you in terms of like, oh, there's a problem, this is how I might troubleshoot it, because I know the way things should be. If you don't have that variety of knowledge to pull from, you're more limited in the ways that you can think about the ways it might fail.'' }
\end{quote}
\change{Developers form rich mental models of how software systems work and how they break.} More years and breadth of experience likely correlates with a more developed intuitive sense.

\rev{In Figure~\ref{fig:experiential_intuition}, \change{we distinguish} between two kinds of knowledge that contribute to experiential intuition: \change{\textit{Specific knowledge}} comes from familiarity with a specific area of code.  \change{\textit{General knowledge}} comes from familiarity with different types of software systems and engineering skills.  P15 describes how this familiarity shapes troubleshooting difficulty:}
\begin{quote}
\textit{``Well, it's easy if you have intimate knowledge of the code that is generating the error, or if you've run up against the error several times before. It's your experience that you have, makes it easy. What's hard is if you encounter a new error that you've never seen before, if it's coming from a different service that you don't own, you don't know a lot about, or a different part of the code that you don't know much about.''}
\end{quote}
\rev{If familiarity has such a significant impact on troubleshooting difficulty, then having an experienced team that maintains familiarity across parts of the system may be a critical factor to keeping the difficulties of troubleshooting under control.}

\subsection{Figuring It Out} \label{figure_it_out}

\rev{At the end of the troubleshooting arc is a pivotal transition—when confusion lifts and the developer is finally able to make sense of the behavior and produces an updated mental model. This moment of insight, which we call \textit{figuring it out}, marks the resolution phase in our model.} In this section, we share qualitative accounts from developers \rev{that reveal} meaningful nuance in how this moment unfolds.  In \textit{discovering the cause}, we explore how developers describe the moment of insight when the cause becomes clear and the developer produces an updated mental model. In \textit{feelings of relief, and sometimes feeling good, and sometimes feeling even more frustrated}, we highlight the complex emotions that can arise after solving the problem and the contextual factors that shape the experience.

\subsubsection{Discovering the Cause}

\textit{Figuring it out} occurs when the developer discovers a key insight that explains the cause of the unexpected behavior. This is the moment when coherence is restored: the system’s behavior makes sense again, and the developer’s understanding re-aligns with what they observe. P14 described this moment of clarity:
\begin{quote}
\textit{``And it was sort of obvious once I just took one look. It was like, oh yeah, there's actually no file-based source of truth for this. So that's why.''}
\end{quote}
\rev{When the developer figures it out, they often articulate their new understanding of why the system output now makes sense. The output of the overall troubleshooting process is an updated mental model and the ability to explain the behavior. In the broader debugging activity (Figure~\ref{fig:trouble_debug}), this new understanding becomes the input to fixing the behavior.}

\subsubsection{Feelings of Relief, and sometimes Feeling Good, and sometimes feeling even more Frustrated} \label{feelings}

After overcoming the confusion, developers describe a variety of feelings that are sometimes mixed and complex.  The most commonly reported feelings are relief and feeling good. P10 described their feelings at the conclusion of troubleshooting:
\begin{quote}
   \textit{``I felt relieved. It was definitely a feeling of success, like I figured it out.''} 
\end{quote}
\change{P02 echoed this feeling of relief, adding nuance to clarify the nature of the feeling:}
\begin{quote}
\textit{``Relieved, I would say. Not like joy or happiness, just like, yes, this is done. Like you have a term paper or something, you just turn that in, and you're like, okay, or a final in school, you hand that in, you're like, okay, I don't know how I did. I don't know if it works, but that's done.''}
\end{quote}
\change{P02's account} highlights another fundamental aspect of developer experience: \textit{uncertainty}. \change{Even after arriving at what seems like a clear understanding, there is always a chance of an incorrect assumption, no matter how confident the developer might be.} 

\change{Several developers reported positive feelings—feeling good, productive, capable, or proud of themselves for overcoming the challenge.  P07 describes what a positive troubleshooting experience feels like: 
\begin{quote}
\textit{``In the end, if everything goes fine, I feel productive. I feel good. I feel like I've done something.''}
\end{quote}
P15 highlighted that it is the good feelings that come with figuring it out that drive the process forward and create intrinsic motivation.  P08 described feeling a dopamine hit at the end, and up to that point feeling frustration.}  

\change{Other developers felt lingering frustration even after the problem was resolved. P10, P14, and P17 felt frustrated when the time spent on troubleshooting seemed preventable. P22 felt like they were ``just kicking the can down the road'' and not making the situation better. } P14 described the complexity of their feelings:
\begin{quote}
\textit{``I mean, proud of myself, but also just still frustrated at the people for having left the problem in the first place, right? It's like a fleeting congratulations. You figured out what was confusing you, but still, it's like why did they leave it in a stupid state?  I would say it's like, you go all the way to very frustrated, and then you come back just a little bit.''}
\end{quote}
\change{The variety of accounts suggest that the emotional experience of troubleshooting is shaped not only by resolving the immediate problem, but also by broader reflections—on team dynamics, time spent, and whether the work felt meaningful or wasteful. Even when the problem is resolved, frustration may persist if the situation evokes a sense of avoidable struggle or systemic neglect, contributing to long-term challenges with morale.  Overall, our theory illuminates the developer's experience of overcoming confusion—revealing the nature of frustration while striving to gain clarity, and the richness and nuance that shapes the experience. }

\section{Discussion} \label{discussion}

\rev{A central aim of our Theory of Troubleshooting is to help developers explain the challenges they face—the confusion, frustration, and loss of control that arises when software systems become difficult to understand. By giving shape to developers’ internal experiences—how it feels to be confused and struggling to gain clarity in a system that makes it difficult to see what is happening—our theory makes these pain points easier to explain. By illuminating the developer's internal experience of troubleshooting, our theory reveals hidden costs and sustainability risks at the heart of software development.}

\rev{In this section, we explore the broader implications of our theory for both research and industry. We begin with \textit{troubleshooting and cognitive fatigue}, and how our theory explains the depletion of cognitive resources that can contribute to fatigue, stress, and burnout. In \textit{sustainability risk and the loss of control}, we discuss project-level implications and the long-term risks associated with rising troubleshooting difficulty. Next, we consider \textit{implications for designing tools} that support developers and \textit{designing software for easier troubleshooting}. We then turn to \textit{resonant language as a conceptual contribution}, sharing feedback from developers on how the theory supports reflection, mentorship, and communication. Finally, in \textit{a new lens for developer experience}, we explore practical applications for researchers and open new possibilities for rethinking productivity in software development.}

\subsection{Troubleshooting and Cognitive Fatigue}

By surfacing and naming the \textit{confusion experience} in terms of neurological and attentional dynamics, our theory provides a conceptual foundation for understanding how prolonged troubleshooting can deplete cognitive resources and lead to fatigue. The puzzling event that initiates the confusion experience triggers an \textit{orientation response} in the body—\rev{an involuntary pull of attention toward understanding the anomaly, and a boost in attentional resources to support hyper-attentiveness \cite{kahneman_attention_1973}.}  This moment also appears to correspond with activation of the insula, a brain region linked to attention control and decision-making under uncertainty—suggesting a neural correlate to the felt experience of surprise and reorientation \cite{duraes_wap_2016}.

\rev{As troubleshooting continues with the confusion unresolved, attentional resources are depleted more rapidly. Developers described increasing frustration, the need to take breaks, and ``blindness effects'—such as not seeing something, or seeing the opposite—symptoms that mirror known effects of cognitive fatigue \cite{boksem_effects_2005}.} When a developer exhausts their attentional capacity (Figure \ref{fig:kahneman_attention}), \rev{they lose their ability to focus on the task at the level the task demands}, and experience cognitive fatigue, the urge to stop, and the need to rest to restore capacity \cite{ackerman_cognitive_2011, kahneman_attention_1973}.

\rev{Cognitive fatigue research suggests that a person may try to push through this exhaustion by expending compensatory effort—forcing themselves to focus despite dwindling resources \cite{ackerman_cognitive_2011}. This coping strategy adds another layer to the fatigue dynamic: while such effort can preserve short-term performance, it comes at a physiological cost of increased stress hormones (e.g. cortisol, noradrenalin), elevated blood pressure, and an even stronger experience of fatigue \cite{robert_j_hockey_compensatory_1997}. When software teams operate under high pressure and there are consequences for a deficit in performance, developers may feel compelled to push through the fatigue—contributing to the risks of burnout and excessive stress.} 

\rev{By tracing how prolonged confusion depletes cognitive resources and leads to fatigue, our theory reframes the mental strain developers are experiencing as a \textit{capacity constraint}, rather than simply an emotion.}

\subsection{Sustainability Risk and the Loss of Control}

As software systems evolve, they often grow in complexity—accumulating dependencies, behaviors, and design decisions that can become harder to reason about over time.  While much attention has been paid to the problem of technical debt \cite{li2015systematic}, our findings suggest that an \rev{overlooked hidden cost and long-term risk} is the rising cognitive cost of troubleshooting.  Maintaining a system requires that developers can make sense of unexpected behavior when it occurs. If the time it takes to troubleshoot increases, it may indicate not just a temporary difficulty, but a deeper loss in the team's ability to sustain understanding of the system.

\rev{Our theory highlights} several ways this loss of sustainability can unfold. For example, developers rely on \textit{experiential intuition} to guide their sense of what to investigate—an intuition built from prior encounters with the system, knowledge shared among teammates, and a career's worth of experience. When familiarity is lost due to team turnover, developers may find it harder to think of things to try that could generate breakthrough insights, leading to an increase in the time spent troubleshooting. Likewise, successful troubleshooting often hinges on the developer’s ability to \textit{poke and see}--by running experiments that generate clues of partial information, the developer can gain clarity and reestablish understanding of how the system works.  When observability is limited, or the runtime behavior is hard to trigger in isolation or with the needed inputs, it can be difficult to uncover new clues.  These experiential difficulties—arising from the interactions between the developer and the system—are central to the developer's felt experience of maintainability, even though they are not direct properties of the code.  As troubleshooting becomes more difficult, developers spend more time confused and unable to make sense of the system.  Over time, this rising troubleshooting effort can become a leading indicator of risk—signaling when the team's ability to sustain understanding is beginning to erode.

\begin{figure*}[h]
\centerline{\includegraphics[scale=0.45]{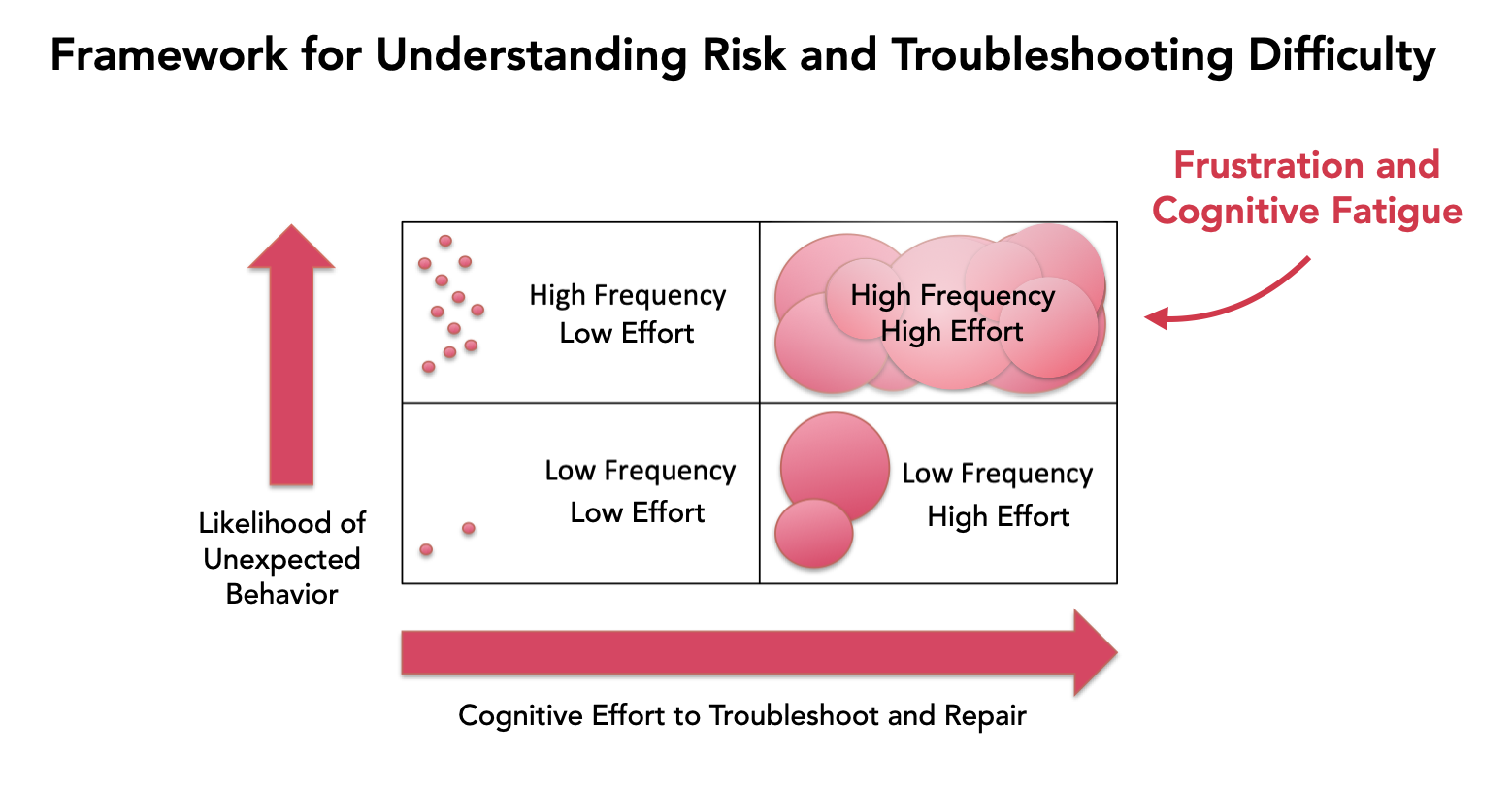}}
  \caption{Synthesis of findings into a simple 2x2 framework illustrating the relationship between the likelihood of unexpected behavior (puzzling events) and the cognitive effort required to troubleshoot, highlighting how compounding difficulties can lead to frustration and cognitive fatigue}
  \label{fig:2x2}
\end{figure*}

This reframes how we understand \rev{long-term project risks in software development}. \textbf{Troubleshooting time becomes a leading indicator of sustainability risk and loss of control in software systems—because it reflects the developer’s ability to sustain understanding amid increasing complexity.} Figure~\ref{fig:2x2} synthesizes our findings into a simple 2x2 framework illustrating the relationship between the likelihood of unexpected behavior (puzzling events) and the cognitive effort required to troubleshoot, highlighting how compounding difficulties can lead to cognitive fatigue.

\change{Where the technical debt metaphor invites teams to locate problems in specific pieces of code, our findings suggest a broader, experience-oriented perspective: what matters is not just the state of the code, but how hard it is for developers to make sense of what the system is doing. Teams may spend significant effort cleaning up code that looks problematic, yet continue to struggle with the same kinds of troubleshooting difficulties if the factors affecting experience are not addressed. By examining actual troubleshooting experiences and identifying what made them difficult to resolve, teams can better target improvements that reduce real friction.}

\subsection{Implications for Designing Tools}

If troubleshooting time is a leading indicator of sustainability risk, then being able to observe and reflect on these experiences becomes crucial. Yet most software tools today offer little visibility into when developers are confused and stuck, what makes troubleshooting difficult, or how long it takes to reestablish understanding. These moments are rarely captured in notes or logs, and the underlying experience—what it feels like to be stuck and working toward clarity—is internal, often tacit, and may not be recognized by the developer as it happens.

\change{This gap points to an opportunity: to develop tools that help make troubleshooting experiences more visible, reflectable, and discussable. Capturing the troubleshooting session—through lightweight annotations, journaling, or structured notes—could support the developer in talking it out, while also creating a record of the experience for individual reflection and team learning. Tools that surface the developer’s experience of confusion and gathering clues, could open new possibilities for understanding and improving Developer Experience.}

\rev{When software teams are identifying the highest-impact problems to solve, tools could support this activity by highlighting where developers are experiencing actual troubleshooting difficulty, rather than assuming problems based on static code analysis. Tools that capture developer troubleshooting sessions could correlate those sessions with the area of code under investigation, helping to automate data collection that supports team discussions. As teams identify troubleshooting hotspots and gain clarity on what makes them difficult, tools could enable the developers to enrich the data so it reflects the team's emergent understanding.  Teams could explore the data together, discuss the challenges they see, and design improvements based on a grounded analysis of the developers' experiences. }

\subsection{Designing Software for Easier Troubleshooting}

\rev{We propose that designing software systems for easier troubleshooting, and investing in improvements that make troubleshooting easier, are a practical strategy for managing the risk of cognitive fatigue.} Our theory helps developers identify improvements that are likely to have an impact. By studying specific examples of long troubleshooting times, we can unpack the multi-factor nature of difficulty and how those difficulties might persist in future troubleshooting. What methods and surfaces did the developer have available to poke at the system and see what was happening?  How did the developer go about generating clues of partial information to gain clarity?  How did the developer's experiential intuition support or hinder their ability to think of things to try?  The \rev{theory offers} a rich set of questions to ask that help reveal the nuanced dynamics of troubleshooting difficulty, and reveal patterns and insights that might otherwise be overlooked.

\change{As a cautionary contrast, recent studies suggest that AI-assisted coding tools intended to improve productivity may be inadvertently increasing the difficulty of troubleshooting \cite{vaithilingam_expectation_2022, barke_grounded_2023}.  In Section~\ref{lab}, we discussed two recent studies by Vaithilingam et al. (n=24) \cite{vaithilingam_expectation_2022} and Barke et al. (n=20) \cite{barke_grounded_2023}} that investigate the lived experiences of developers using GitHub Copilot, and both studies show a pattern where developers had an easier time getting started with the AI tools, but then faced increased difficulties when troubleshooting the generated code.  Vaithilingam's study compared developers' ability to complete the same task, with or without Copilot, and found a slight improvement in the overall task-completion time with Copilot (1 min), but three of the developers using Copilot failed to complete the task due to a time-consuming debugging effort, and an underestimation of the time it would take to fix the bug \cite{vaithilingam_expectation_2022}.  All 12 of the developers that attempted to repair the generated code reported finding it more difficult to repair the code they did not write themselves.  P7 from Vaithilingam's study said:
\begin{quote}
\textit{``It made debugging the code more difficult as I hadn’t written the code directly and didn’t have an initial intuition about where the bugs might be. Especially with a final bug in my program I really had no idea why it was happening and had to refactor the code.'' \cite{vaithilingam_expectation_2022}}
\end{quote}
The intuition that P7 describes is the same \textit{experiential intuition} we described in our theory that developers rely on to guide their strategy and decide what things to try.  Typically when writing the code (without Copilot), a developer would deeply engage with understanding the program, then iteratively think through and construct the logic of the new code changes line-by-line, resulting in both the code output, as well as a deep familiarity of the specific code that was just written.  Upon running the code and discovering an error, the developer's experiential intuition gives an immediate hunch as to what the problem is, making debugging fairly quick.  When Copilot generates the code, the code output is there, but this deep familiarity of the code that influences experiential intuition is missing.  

\change{Our theory explains why tools intended to improve productivity may paradoxically shift the burden toward more mentally-taxing, cognitively fatiguing, and potentially frustrating work.}  In terms of Kahneman's model, \change{if developers generate the code faster but then spend more time troubleshooting,} the result is a non-linear trade-off of attentional resource consumption.  If organizations only evaluate short-term productivity, and ignore the impact of cognitive fatigue, there's a risk of latent problems building up over time with increasing mistakes and burnout.  Further, the long-term impact of developers not building a deep familiarity of the code to support their experiential intuition, could have unforeseen consequences as well.  Understanding these trade-offs through the lens of our \rev{theory} can help teams recognize when well-intentioned improvements introduce new problems, identify the dynamics at play, then design \rev{their software systems} more effectively for easier troubleshooting.

\subsection{Resonant Language as a Conceptual Contribution}

\rev{Our goal in constructing an experience-centered theory was to give structure and language to developers’ internal experiences—often tacit, intuitive, and hard to explain. As P20 noted, \textit{``It’s interesting and difficult to try and put words to these things that I just kind of do intuitively.''} By offering resonant language grounded in developers’ own words, we surface existing tacit knowledge and create conceptual tools that are already attuned to how developers think—making them easier to adopt and use in practice.}

From an HCI perspective, this goal aligns with a tradition of experience-centered research that seeks to illuminate and improve how humans interact with complex systems and tools. Theoretical contributions in this tradition are evaluated not only by their analytical power, but by whether they offer people new ways to see and talk about their experience. In this section, we highlight how participants responded to \rev{our theory}, and explore how resonant language can support communication, mentorship, and shared learning.

\subsubsection{Mentorship and Resonant Language} \label{mentors}
When we sent out our paper to participants to get feedback, one of the most compelling responses came from P25 who shared how the language helped them articulate aspects of their own troubleshooting experience that had previously been difficult to explain—especially in mentorship contexts:

\begin{quote}
\change{\textit{``I managed to carve out some time to read the paper, and I really enjoyed it! Something that really stood out for me was the point you made where developers often have difficulty describing their process—and I immediately thought about how that can be a barrier in teaching/mentorship (I think every senior has had that experience of having a junior teammate just baffled at how you figured something out, and you have trouble explaining exactly what you did!). I really loved the categorization and naming of the various troubleshooting phases and steps, and feel certain that I will be able to use this language to describe my process when I am mentoring.''}}
\end{quote}

\change{This feedback illustrates the generative potential of resonant language—not just as a way to analyze experience, but as a way to talk about it, share it, and learn from it together. In the tradition of HCI research that values theoretical contributions grounded in lived experience, we see this kind of resonance as a signal of conceptual utility: the language helped make something tacit more visible, and gave practitioners new ways to reflect on and communicate their process.}

In other responses, participants shared how the paper helped them reflect on their own experiences and sparked new thinking. \rev{P18} noted how the paper made them consider the role of environments that encourage sharing frustrations and learning from failure. \rev{P21} described how the language and distinctions helped them reframe their thinking about troubleshooting, context switching, and the challenges of working in legacy systems. Collectively, this feedback points to the conceptual utility of the language: it not only resonated with the lived experience of developers, but also supported new insight, reflection, and practical application.

\rev{The model diagrams also functioned individually as condensed representations of the theoretical constructs and were strongly resonant with developers. During the follow-up interviews, we invited participants to respond to each diagram and reflect on how well it aligned with their experiences. Many participants ``tried on' the language by reflecting on their own experiences using the models. In some cases, they adopted the language immediately, sharing new insights about their experiences revealed by the model.  These interactions not only confirmed the resonance and clarity of the visuals, but also demonstrated how the diagrams functioned as usable sense-making tools.}

\subsubsection{Communicating Troubleshooting Difficulties}

To gain support for spending time on improvements, developers need to communicate the difficulties \rev{they experience} to their managers.  \rev{Our theory offers a way for developers to explain their experiences as the software becomes increasingly complex, without having to reach for distant metaphors like technical debt.} For managers that may not have direct experience of what software development is like, the experience of confusion creates a common reference point to understand what is happening.  Both developers and managers can likely relate to what it feels like to be confused and stuck for hours, and how frustrating and cognitively fatiguing that experience can be.  As the software becomes increasingly complex, we can imagine encountering more circumstances of puzzling events and confusion, and it becoming increasingly difficult to troubleshoot.  With a shared understanding of the problems that is closer to the felt experience and phenomena being encountered, our \rev{theory helps} to facilitate communication of the difficulties, and get managers and developers on the same page.

Whether mentoring teammates, advocating for improvements, or reflecting on experience, resonant language offers general utility that reinforces its value as a conceptual contribution. By surfacing nuanced aspects of troubleshooting, our \rev{theory provides} more than just analytical insight—\rev{it offers} language that developers can use to explain, teach, and advocate for change in their tools and environments.  In experience-centered HCI research, this kind of conceptual clarity and communicative power is a hallmark of meaningful theoretical contribution.

\subsection{A New Lens for Developer Experience}

\rev{In this final section, we discuss two implications of our theory. First, we discuss the expanded lens of Developer Experience, and offer practical applications for researchers conducting observational studies. Then, we return to the foundational questions raised in the introduction about the communication gap between developers and managers and the current focus on productivity.}

\subsubsection{Observing Developer Experience over Time}

\rev{Theoretical constructs such as \textit{puzzling events}, the \textit{confusion experience}, and \textit{experiential intuition} offer researchers a new lens for studying developer behavior and reorienting how we think about Developer Experience. Our theory \rev{describes} thinking, feeling, and striving over time, offering a temporal perspective that helps surface the nuanced and multi-factor influences shaping developers’ internal experiences. This framing applies beyond just troubleshooting—it is a foundational paradigm for understanding the developer's lived experience as it unfolds over time.}

\rev{For researchers, the theoretical constructs offer an experience-centered coding vocabulary for observing real troubleshooting sessions.  For example, researchers could identify puzzling events, track how the confusion experience unfolds, and observe how developers generate clues of partial information to gain clarity. Similarly, patterns in developer behavior—what they notice or try—could be studied through the lens of experiential intuition.  By learning to recognize how developers' internal experiences map to observable behavior, studies can develop coding schemes that deepen our understanding of Developer Experience.}

\subsubsection{Reframing the Goal of Productivity: Sustainable, Joyful, and Effective at the Same Time}

\rev{Productivity has long been the default lens through which both research and industry evaluate software development. While this lens does capture aspects of performance, it also conceptually flattens the deeper meaning of what we are aiming for—making it difficult to have more nuanced conversations about how we could support work that is sustainable, joyful, and effective, at the same time. Returning to where we began this journey, Hans Dockter’s keynote at the first DPE Summit \cite{gradle_dpe_2022} highlighted a core belief of ``joy == productivity,' expressing how joy and productivity are actually in alignment with one another: with smart investments in engineering capabilities, it is possible to unlock innovation capability, productivity, and joyful developer experiences at the same time.}

\rev{One critical dimension that a narrow focus on productivity tends to overlook is sustainability. Our findings suggest that rising troubleshooting effort may serve as a high-signal indicator of sustainability risk—revealing when a team’s ability to understand and maintain a system is beginning to erode. Yet today, this dimension is largely absent from Developer Experience frameworks and conversations. Our research highlights the importance of bringing this perspective in focus.}

\rev{The challenge Dockter identified was a communication breakdown between developers and managers. His insight was that cognitive fatigue could be a conceptual bridge—one that could make the developers’ intuitive sense of strain, confusion, and loss of control more legible to others.  Over time, as productivity has become the dominant framing for evaluating software development, these foundational insights have faded from view—but they remain vital to addressing the deeper challenges organizations face. Without shared conceptual foundations, organizations struggle to recognize what is breaking down, why it matters, and where investment would truly make a difference.  Our theory offers a conceptual foundation rooted in cognitive science that helps developers explain the challenges they face and the risks that emerge when troubleshooting becomes difficult—without relying on metaphors like technical debt. By clarifying how prolonged troubleshooting drains cognitive resources and contributes to fatigue, our theory enables managers to recognize the project risks and understand the significance of investing in improvements.}

\rev{This theory opens a new possibility. When we center the developer’s lived experience—not just their output, but how it feels to do the work—we open up new possibilities. What if software development could be redesigned to support both high productivity \textit{and} a sense of energy, flow, and joy? Psychology research shows that, under the right conditions, cognitively demanding work does not have to be fatiguing. When a task is self-initiated, intrinsically motivated, and especially when the activity is regarded as play—cognitively demanding work can be sustained for long periods without leading to fatigue \cite{ackerman_cognitive_2011}. Instead, such work may generate feelings of energy, alertness, and even elation \cite{csikszentmihalyi1990flow}. Hockey and Earle further identified \textit{agency} in the ability to choose one’s tasks and schedule—as a key factor in whether significant fatigue occurs \cite{hockey_control_2006}. This flow-state effect that protects against fatigue is a rich area for future research. We invite researchers, practitioners, and entrepreneurs to explore how development environments, team practices, and organizational structures might be reimagined to support this vision of sustainable, enjoyable, and high-performing teams.}

\subsection{Evaluating the Theoretical Contribution} \label{evaluate}

\change{In closing, we reflect on the contribution of this study using evaluation criteria from Constructivist Grounded Theory \cite{charmaz_constructing_2014}. These criteria—credibility, originality, resonance, and usefulness—are particularly well-suited for evaluating theories grounded in lived experience and aimed at meaning-making in complex practice.}

\vspace{0.5em}
\textbf{Credibility:} \rev{The study draws from 27 interviews with professional developers, representing 570 cumulative years of experience, providing a rich dataset for analysis. The theory was developed using a rigorous process of initial grounded coding with high-resolution descriptive codes, constant comparison, and memoing to construct the emerging theory.}

\vspace{0.5em}
\textbf{Originality:} The Theory of Troubleshooting offers a novel synthesis of thinking, feeling, and striving dimensions in temporal models \rev{of the developer’s cognitive experience.} It surfaces under-recognized patterns such as the confusion experience and experiential intuition, and introduces resonant language to articulate these dynamics. \rev{By connecting these insights to cognitive science, the theory offers a new lens for understanding Developer Experience.}

\vspace{0.5em}
\textbf{Resonance:} Seven follow-up interviews were conducted to evaluate resonance, gather feedback, and refine the models. Participants indicated that the models resonated with them and meaningfully captured aspects of their own experience. Feedback on the paper indicated that the terminology was relatable, practically useful, and supported reflection and communication.

\vspace{0.5em}
\textbf{Usefulness:} \rev{Our theory highlights connections between troubleshooting, confusion, and cognitive fatigue, and offers new leverage points for improving developer tools, mentoring practices, and the long-term sustainability of software systems. By making the developer's experience of troubleshooting more legible, our theory helps improve communication between developers and managers, and support team discussions and improvement efforts.}

\vspace{0.5em}
\change{Taken together, these criteria affirm the strength and grounding of the theory. At the same time, we recognize that every theory is partial, situated, and evolving. In the following sections, we reflect on the study’s limitations and identify opportunities for future work.}

\subsection{Future Work}

\rev{One of our key motivations for developing this theory was to bridge the communication gap between developers and managers, yet in this study we focused on making the developers experiences more legible.  While we validated that the theory resonates with developers, we have yet to test whether our theory bridges the communication gap in practice. Future work could investigate how these constructs function in real conversations between developers and managers. Can developers use this language to express their experiences? Are managers able to get a better idea of what the developers' experiences feel like, and be more in alignment on decisions as a result?  These questions are essential to testing the practical utility of this theory.}

\rev{Our dataset also reveals rich dynamics of team-based troubleshooting—how developers collaborate to get unstuck, and the cultural challenges that can deter collaboration under organizational pressure to perform as individuals. While team-based troubleshooting is beyond the scope of this paper, it offers fertile ground for future theory development.}

\rev{Building on this theory, there is an opportunity to design and evaluate interventions with tools. For example, future work could design a prototype that captures real troubleshooting sessions, correlates them with areas of code, and highlights segments where developers spend significant time confused. These tools could help surface difficulty hotspots, and inform improvement decisions in real-world settings. Future research could use a design science approach to investigate and evaluate the interventions.}

\section{Conclusion}

\change{This study presents a Theory of Troubleshooting grounded in the lived experiences of 27 professional software developers. Using Constructivist Grounded Theory, we constructed a set of experience-centered models that capture the developer's cognitive experience of overcoming confusion. \rev{Key theoretical constructs} include the \textit{confusion experience}, \textit{experiential intuition}, and \textit{poking and seeing}—each \rev{illuminating} a different facet of the internal experience.}

\change{A central contribution of this theory is the insight it offers into the developer's internal experience of troubleshooting. By modeling the thinking, feeling and striving of the developer over time, \rev{our theory contributes} a process-oriented lens for Developer Experience.  We highlight the importance of unflattening concepts like ``productivity' so we can make room for the nuance and richness to express what we truly aim for.}  

\change{One critical dimension that a narrow focus on productivity tends to overlook is sustainability. Our findings suggest that rising troubleshooting effort may serve as a high-signal indicator of sustainability risk—revealing when a team’s ability to understand and maintain a system is beginning to erode. We synthesized our findings into a simple 2x2 framework illustrating the relationship between the likelihood of unexpected behavior and the cognitive effort required to troubleshoot, highlighting how compounding difficulties can lead to cognitive fatigue.  We propose that designing software systems for easier troubleshooting, and investing in improvements that make troubleshooting easier, are a practical strategy to reduce the risk of cognitive fatigue.}

These insights open new possibilities for practice, reflection, and collaboration.  Developers might use these concepts to explain their struggles more clearly, reflect on patterns of difficulty, or surface hotspots in the system. For researchers, these constructs offer an experience-centered coding vocabulary for observing real troubleshooting sessions.  For managers or tool designers, our theory provides a communication bridge—making the invisible dynamics of troubleshooting more legible, and creating space for more effective collaboration.

\section{Acknowledgments}

\rev{We would like to thank Todd Sedano for providing feedback and guidance on our methodological approach with CGT, the participants in the study for their valuable contributions, and Norman Anderson, Umit Akirmak, Nathan Cassee, Keith Mann, and the reviewers for their time, engagement, and thoughtful feedback that made this manuscript better.}

\bibliographystyle{ACM-Reference-Format}
\bibliography{references}

\pagebreak

\section{Appendix} \label{appendix}

\change{Included in the Appendix are 1) the emerging questions we explored before narrowing our theoretical focus including links to Miro boards with 1032 initial grounded codes sorted by question, 2) a detailed demographics table, 3) \rev{a summary and }detailed counts table showing the contribution of data across all 27 interviews to each theoretical category, 4) a conceptual diagram of data analysis terminology 5) a list of the 73 theoretical categories that emerged from the data analysis before narrowing our theoretical focus, and 6) the interview protocols for the first and second (follow-up) rounds of interviews. In the follow-up interview protocol, we show an early revision of the model diagrams before any feedback was received to demonstrate how the models evolved during the study and show our work.}

We also provide a data archive on Zenodo (\underline{\url{https://zenodo.org/records/14948399}}) which includes the eight theoretical category reports with 681 initial grounded codes across all 27 interviews.  \rev{If a researcher wants to understand the variety of ways the \textit{confusion experience} can arise in developer experience, they can consult a dedicated report that includes 119 initial grounded codes, each labeled with corresponding participant numbers.  We have also made available in the archive the 16 emerging question reports of 1032 grounded codes that were used to populate the Miro boards. These materials can help researchers deepen their understanding of the theoretical constructs when using the theory for research coding in their own studies.}

\pagebreak

\subsection{Emerging Questions and Miro Boards} \label{rq}
\vspace{1em}
\noindent\begin{minipage}{\linewidth}
\textbf{Developer Context Questions:}

\begin{itemize}
\item {What kind of role is the developer currently working in? 
 \href{https://miro.com/app/board/uXjVNoWzlTM=/}{\underline{Miro Board}}}
\item {What type of development environments are developers working in?  Are they more sophisticated/mature with infrastructure automation, or more legacy?  
 \href{https://miro.com/app/board/uXjVNoXTIMQ=/}{\underline{Miro Board}}
}
\item {What type of company environment is the developer working in?  Is it a startup or big corporate?  What’s the company and team culture like?
 \href{https://miro.com/app/board/uXjVNoaoPf0=/}{\underline{Miro Board}}
}
\item {What is the organizational/team structure that the developer works in?  Who does the developer collaborate with?
 \href{https://miro.com/app/board/uXjVNoapAFE=/}{\underline{Miro Board}}
}
\item {How do developers identify themselves?
 \href{https://miro.com/app/board/uXjVNoXXF4s=/}{\underline{Miro Board}}
}
\end{itemize}
\vspace{1em}
\textbf{Troubleshooting Questions:}

\begin{itemize}
\item {What are the different types of troubleshooting situations people see?
 \href{https://miro.com/app/board/uXjVNlo9SCY=/}{\underline{Miro Board}}
}
\item {What are the different steps/stages/subprocesses involved in the troubleshooting process?
 \href{https://miro.com/app/board/uXjVNk7LhFA=/}{\underline{Miro Board}}
}
\item {What are the different strategies developers use to troubleshoot?
 \href{https://miro.com/app/board/uXjVNlbpgaY=/}{\underline{Miro Board}}}
\item {What makes troubleshooting easy vs difficult?
 \href{https://miro.com/app/board/uXjVNmfK2fA=/}{\underline{Miro Board}}
}
\item {How do developers feel while they are troubleshooting?  How do their feelings change before and after?  What factors influence how they feel?
 \href{https://miro.com/app/board/uXjVNmfpj-U=/}{\underline{Miro Board}}
}
\item {How do developers learn how to troubleshoot? (emergent)
 \href{https://miro.com/app/board/uXjVNk4mmno=/}{\underline{Miro Board}}}
\end{itemize}
\vspace{1em}
\textbf{Asking for Help Questions:}

\begin{itemize}
\item {How do developers reach out for help?  Who do they decide to ask and why?
 \href{https://miro.com/app/board/uXjVNk47nic=/}{\underline{Miro Board}}
}
\item {Does the developer lean toward collaborating or anti-ask for help?  What do these different cultural norms look like around collaboration?
 \href{https://miro.com/app/board/uXjVNk_nbxw=/}{\underline{Miro Board}}
}
\item {Why do developers decide not to ask for help?
 \href{https://miro.com/app/board/uXjVNk4449c=/}{\underline{Miro Board}}
}
\item {How do developers collaborate and try to help one another?  What does the process look like?  What are the strategies developers use to collaborate?
 \href{https://miro.com/app/board/uXjVNk9a2m0=/}{\underline{Miro Board}}}
\item {Is collaborating effective?  What makes it effective?
 \href{https://miro.com/app/board/uXjVNk95g-s=/}{\underline{Miro Board}}
}
\end{itemize}
\end{minipage}

\subsection{Demographic Information By Participant}

\begin{center}
\begin{tabular}{l l r l l r l}
\hline
Participant & Organization Size & Team Size & Remote/Colocated/Hybrid & Education & YearsExperience & Location \\
\hline
P01 & \textgreater{} 1000 people & 30 & Remote & Bachelor's degree & 42 & US \\
P02 & \textgreater{} 1000 people & 12 & Remote & Bachelor's degree & 8 & US \\
P03 & \textgreater{} 1000 people & 8 & Remote & Post-graduate degree & 15 & US \\
P04 & 1-50 people & 1 & Remote & Bachelor's degree & 35 & US \\
P05 & 51-200 people & 10 & Remote & Bachelor's degree & 20 & US \\
P06 & 201-1000 people & 3 & Hybrid work & (Not reported) & 15 & Europe \\
P07 & \textgreater{} 1000 people & 6 & Remote & Post-graduate degree & 5 & US \\
P08 & \textgreater{} 1000 people & 3 & Remote & Bachelor's degree & 34 & US \\
P09 & 1-50 people & 1 & Remote & Post-graduate degree & 28 & US \\
P10 & \textgreater{} 1000 people & 4 & Colocated & Some college, no degree & 25 & US \\
P11 & 51-200 people & 5 & Hybrid work & Post-graduate degree & 8 & Canada \\
P12 & \textgreater{} 1000 people & 10 & Remote & Bachelor's degree & 24 & Canada \\
P13 & \textgreater{} 1000 people & 10 & Remote & Bachelor's degree & 28 & ? \\
P14 & \textgreater{} 1000 people & 8 & Remote & Bachelor's degree & 22 & Europe \\
P15 & \textgreater{} 1000 people & 8 & Remote & Post-graduate degree & 29 & US \\
P16 & 1-50 people & 6 & Remote & Bachelor's degree & 25 & Canada \\
P17 & 51-200 people & 8 & Remote & Bachelor's degree & 9 & US \\
P18 & 51-200 people & 3 & Remote & Bachelor's degree & 28 & Europe \\
P19 & 1-50 people & 10 & Remote & Bachelor's degree & 18 & US \\
P20 & 51-200 people & 16 & Remote & Bachelor's degree & 15 & Canada \\
P21 & 51-200 people & 3 & Remote & Bachelor's degree & 30 & US \\
P22 & \textgreater{} 1000 people & 12 & Remote & Bachelor's degree & 21 & Canada \\
P23 & 51-200 people & 2 & Remote & Post-graduate degree & 18 & US \\
P24 & 201-1000 people & 7 & Remote & Post-graduate degree & 29 & US \\
P25 & 1-50 people & 3 & Remote & Bachelor's degree & 11 & Canada \\
P26 & \textgreater{} 1000 people & 7 & Remote & Post-graduate degree & 12 & Canada \\
P27 & 1-50 people & 3 & Remote & Bachelor's degree & 16 & Canada \\
\hline
\end{tabular}

\vspace{1ex}
\parbox{1.0\linewidth}{
\footnotesize
\textit{Gender information is summarized to preserve participant anonymity. Eight participants were identified as women, one as non-binary, and 18 as men, based on publicly available information on LinkedIn (unconfirmed). }
}
\end{center}

\subsection{Contribution of Grounding Across Theoretical Categories} 

\vspace{1em}

\begin{tabular}{l|c|c}
\hline
\textbf{Theoretical Category} & \textbf{Participants (n)} & \textbf{Initial Grounded Codes} \\
\hline
Confusion Experience         & 26 & 119 \\
Trouble in the Creation Process & 9  & 16  \\
Trying to Gain Clarity       & 23 & 126 \\
Poking and Seeing            & 26 & 142 \\
Elucidating the Problem      & 20 & 48  \\
Frustration vs Confidence    & 23 & 84  \\
Experiential Intuition       & 27 & 114 \\
Figuring It Out              & 20 & 32  \\
\hline
\end{tabular}

\vspace{2em}

\noindent
\begin{tabular}{lcccp{3cm}p{3cm}}
\hline
\textbf{Theoretical Category} & \textbf{Participants} & \textbf{Codes G1} & \textbf{Codes G2} & \textbf{Group 1 Participants} & \textbf{Group 2 Participants} \\
\hline
Confusion Experience & 26 & 41 & 78 & P02, P04, P06, P10, P11, P14, P15, P16, P19, P23, P25, P27 & P01, P03, P05, P07, P08, P09, P12, P13, P17, P20, P21, P22, P24, P26 \\
\hline
Trouble in Creation Process & 9 & 14 & 2 & P04, P10, P14, P15, P16, P23, P25 & P17, P24 \\
\hline
Experiential Intuition & 27 & 54 & 60 & P02, P04, P06, P10, P11, P14, P15, P16, P19, P23, P25, P27 & P01, P03, P05, P07, P08, P09, P12, P13, P17, P18, P20, P21, P22, P24, P26 \\
\hline
Poking and Seeing & 26 & 80 & 62 & P02, P06, P10, P11, P14, P15, P16, P19, P23, P25, P27 & P01, P03, P05, P07, P08, P09, P12, P13, P17, P18, P20, P21, P22, P24, P26 \\
\hline
Trying to Gain Clarity & 23 & 81 & 45 & P02, P06, P10, P11, P14, P15, P16, P19, P23, P25 & P01, P03, P05, P07, P08, P09, P17, P18, P20, P21, P22, P26 \\
\hline
Frustration vs Confidence & 23 & 56 & 28 & P02, P04, P06, P10, P11, P14, P15, P16, P19, P23, P25, P27 & P03, P05, P07, P08, P13, P17, P18, P20, P21, P22, P24 \\
\hline
Elucidating the Problem & 20 & 34 & 14 & P04, P06, P10, P14, P15, P16, P19, P23, P25, P27 & P03, P07, P09, P12, P13, P18, P20, P21, P24, P26 \\
\hline
Figuring It Out & 20 & 12 & 20 & P02, P10, P14, P16, P25 & P01, P03, P05, P07, P08, P09, P12, P13, P17, P21, P22, P24, P26 \\
\hline
\end{tabular}
\vspace{1ex}
\parbox{1.0\linewidth}{
\footnotesize
\textit{Breakdown of theoretical category contributions by participant. Group 1 (G1) includes the first 12 participants used to construct the models prior to reaching theoretical saturation and narrowing focus to eight theoretical categories. Group 2 (G2) includes the remaining 15 participants used to enrich the theoretical categories. }
}



\section*{Supplementary material}

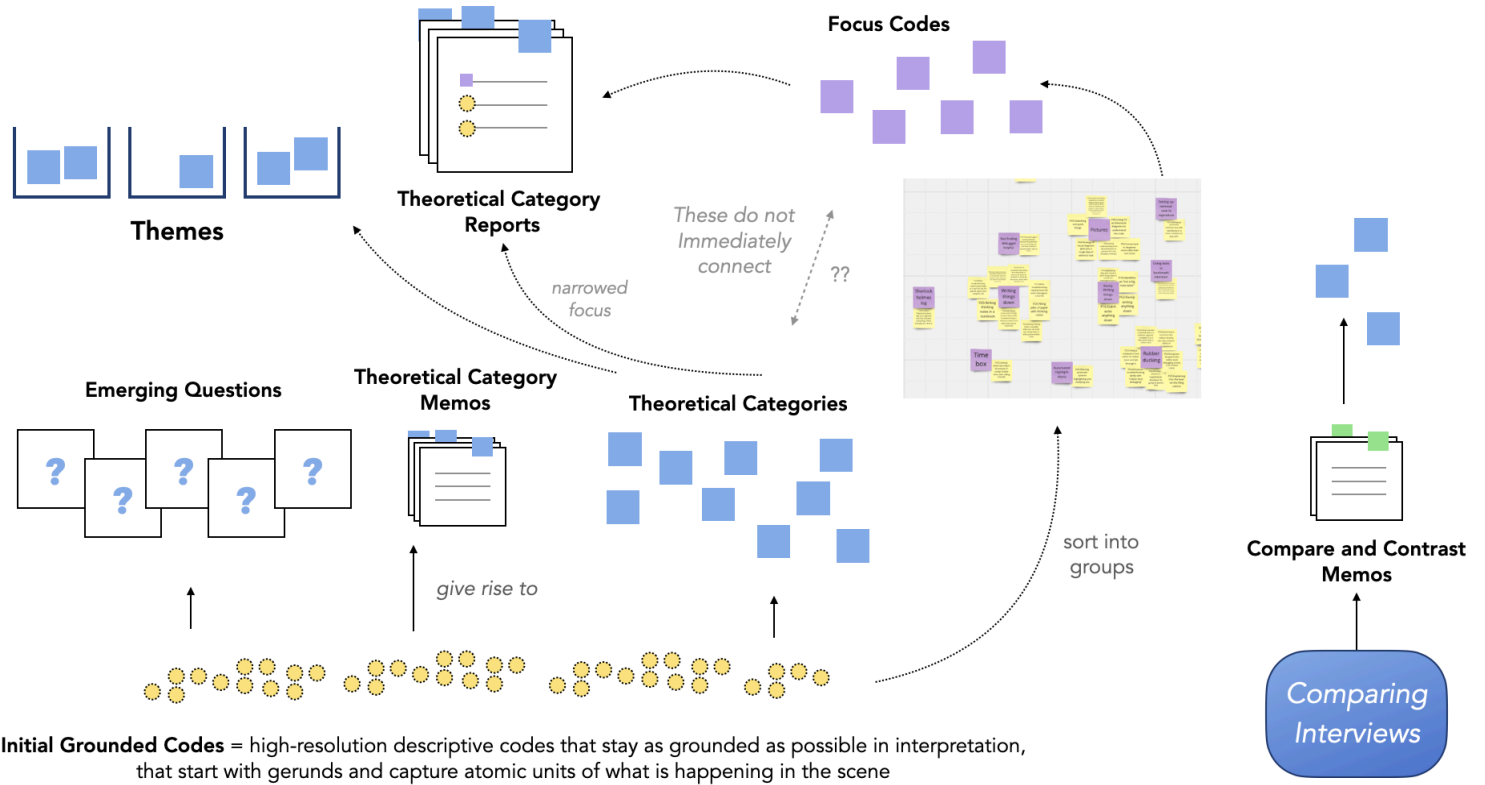

# Grounded Theoretical Categories

**Collaboration**

1. Anti-Ask-For-Help Culture
2. Teaching Juniors
3. Challenge of Giving Context
4. Context Rings
5. Feeling Stretched
6. Coordination Costs
7. Ego Attachment
8. Interruptibility
9. Realtime vs Async
10. Across Timezone Collab
11. Being the Point Person
12. Congeniality on the Team
13. Easy Going vs Anxious
14. Metrics Causing Anti-Help Culture
15. Mob Programming
16. Pop Up a Hand
17. Raising It Up the Flag Pole
18. Respecting the Flow
19. Trivia Team Effect of Multiple Brains
20. Virtual Office Pairing Rooms
21. Code Ownership Database
22. Feeling Lonely and Enjoying Helping
23. Being Good at Crafting Questions

**Experience**

1. Trouble in the Creation Process
2. Frustration vs Confidence
3. Hard to Get Time
4. Steps of Troubleshooting
5. Eureka Insights
6. Notion of Progress
7. Trying to Gain Clarity
8. Confusion Experience
9. Experiential Intuition
10. "Make It Better" Feels
11. Afraid of Breaking Code
12. Building Up Context
13. Curiosity and Wondering
14. Developer Flow
15. Extremely Hard Troubleshooting
16. Impact of Familiarity
17. Mistakes when Shifting Paradigms
18. Monster Complexity
19. New Kid on the Block
20. Running Out of Ideas
21. Unclear Documentation
22. Waiting on Build
23. Wondering Threads
24. Figuring It Out

**Strategy**

1. Observable vs Transparent Systems
2. Rubber Ducking
3. Elucidating the Problem
4. Good Error Messages
5. Bridging Flow to Tomorrow
6. Poking and Seeing
7. Practicing at Troubleshooting
8. Constructing Adhoc Tools
9. Isolating Behavior
10. Observability in Prod
11. Questioning Assumptions
12. Sherlock Holmes Log
13. Staying in Control
14. API Docs Always Open
15. Automating Local Environments
16. Getting Up and Walking Away
17. Learning Up Front
18. Listing Things to Try
19. Making it Safe to Refactor
20. Power of Whiteboarding
21. Book Clubs at Work
22. Demonstrating the Problem
23. Creating a "Repro"

**Thriving**

1. Ability to See What's Happening
2. Empowered to Decide
3. Healthy Teams

# Developer Troubleshooting Study Interview Protocol

**Interview goals:** Collect qualitative data *to understand the experience of troubleshooting in software development, workflows and strategies, difficulties, and collaboration patterns.*

**Introduction:** Interviewer describes the context, the motivation and the goal of the study. The participant is also informed that all information collected will be anonymized and it won't be shared outside the team of researchers directly involved in the project. The interviewer asks the participant for his/her consent to record the interview as well.

*Hi [NAME], and welcome. I am Arty Starr, a researcher from the computer-human interaction and software engineering lab, chisel, at University of Victoria, working in collaboration with my supervisor, Dr. Margaret-Anne Storey.*

*We would like to learn about your experiences of troubleshooting in software development – strategies used, difficulties faced, and how you collaborate with others on your team to troubleshoot problems. My PhD research is focused on developing a theory of momentum in software development, and how we can measure the experience of "friction" and "flow". We are focused on understanding developer troubleshooting, in particular, because it's one of the best opportunities to learn about the cognitive processes of software development, so we can build better tools to support developers.*

*I also want to inform you that I've personally been a software engineer in industry for about 20 years, so it's okay if you speak in familiar developer terms in answering the questions, I'll be able to understand what you mean.*

*Only the researchers involved in this study will see your responses, and we will summarize the responses so they are fully anonymized. Your participation in this study is voluntary, and you do not have to answer any question that makes you uncomfortable, and you can stop the interview whenever you want. We'd like to encourage you to refrain from sharing identifying information as much as possible, such as names of 3rd parties, and to refrain from sharing any confidential information during the interview.*

*The interview should take no longer than 60 min. During the interview we will ask several questions, and we would appreciate it if you respond only based on your personal experiences. If needed, feel free to take your time in answering the questions.*

*So, before we get started, do you have any questions?*

*If this is clear, and you have no more questions, do I have your permission to record the interview?*

**Demographics**

1. Could you please introduce yourself?
2. Could you describe the people you work closely with at your organization?

**Generalized Experiences of Troubleshooting**

We're going to be talking a lot about ***troubleshooting***. If you think about experiences, where you were coding something, then you ran an application or a test, and then got some unexpected output that you

didn't understand, then had to troubleshoot the problem – these friction points that we run into during our everyday coding experiences, is what I want to spend some time understanding better with you.

1. From your experience, what makes these troubleshooting experiences easy versus difficult?
2. What sort of tools and strategies do you normally use, if any, to enhance your ability to troubleshoot?
3. Are there different types of troubleshooting situations?  How would you categorize the different types of situations you encounter?
4. What resources (websites, tutorials, references, communities) do you usually use/access while troubleshooting?  How often do you use them?  Can you give me an example scenario?

**Recalling a Specific Troubleshooting Experience**

1. Can you think of a recent experience, where you were coding something, then you ran an application or a test, and then got some unexpected output that you didn't understand?  First, just try to think of a specific experience.
2. Can you walk me through the experience and steps you went through?  How did you first encounter the problem? (then what happened after that?)
3. Did you write down or record anything while you were troubleshooting?  What sort of tools did you use to help you think through the problem?  (follow up) Did you keep any of this information for future use?
4. How did you feel when you were troubleshooting?
5. How did you feel after you solved the problem?
6. About how long did it take you to troubleshoot the problem?
7. On a scale from 1-10, 1 being the easiest, and 10 being the hardest troubleshooting experience you've ever had, how would you rate this particular troubleshooting experience?


[**Interviewer**: *Just a reminder that all the information provided in this interview will be treated confidentially. Names of companies and people will be anonymized and not shared outside the research team*]

**Generalized Experiences of Team Troubleshooting Collaboration**

Next, let's talk about what happens when we get stuck troubleshooting a hard problem, and need some help from our team.

1. From your experience, when do you typically decide to ask a team member for help?
2. How do you typically reach out?
3. How do you usually decide who to ask?
4. What are some of the challenges that come up with trying to collaborate on a problem?
5. What are some of the strategies you've found that make this collaboration easier?

**Recalling a Specific Ask for Help Experience**

1. What is an experience you remember where you were troubleshooting and asked a team member for help?   First, just try to think of a specific experience.
2. First, what was the problem you were troubleshooting, and what made you decide to ask for help?

3. How did you decide who to ask?
4. How did you go about communicating the problem?
5. How did collaborating affect your ability to resolve the problem?

**Recalling a Specific Helper Experience**

1. What is an experience you remember where someone else on your team was troubleshooting, and you were asked for help? First, just try to think of a specific experience.
2. What were you doing or trying to do when your teammate asked you for help?
3. How did your teammate go about communicating the problem to you?
4. What did you do to help?
5. How did collaborating affect you and your teammate's ability to resolve the problem?

**Ideas for Improvement**

1. What would you suggest are some potential opportunities to support software developers in being more effective at troubleshooting?
2. What would you suggest are some potential opportunities to make team collaborative troubleshooting easier?

**Closing**: Interviewer thanks the participant for joining the study and briefly describes the next steps in the study and the possibility to contact him/her again for a focus group session.

*That was my last question, is there anything else you want to share about your experiences, or things that you think should be included in the study?*

*Thank you for participating in the study, this concludes the interview.*

*For this study, we were interested in learning about your experiences of troubleshooting in software development. We also plan to do a follow up focus group to share what we've learned from the interviews, get feedback, and also demo some ideas for cognitive support tools for software developers, and get your thoughts and feedback on the tooling ideas as well.*

*Would it be alright if we contacted you about participating in the focus group?*

*Our next step is to process and transcribe this interview. We will securely process the recording and the transcript. If at any point in time, you decide you want to withdraw from the study, you can email us and we will remove your data. Also feel free to email us if you have any questions.*

*Do you have any final questions or any further comments?*

*Alright, I'm going to stop recording.*

*Thank you very much for your time.*

# Followup Developer Interview Protocol (Resonance Test)

**Interview goals:** Collect qualitative data *to validate whether the theoretical models constructed during the first set of interviews resonate with developers, and discover whether there are gaps or misunderstandings and how the developers interpret the models.*

**Introduction:** Interviewer describes the context and purpose of the second interview, and recalls the motivation and goal of the study. The participant is also informed that all information collected will be anonymized and it won't be shared outside the team of researchers directly involved in the project. The interviewer asks the participant for his/her consent to record the interview as well.

*Hello [NAME], and welcome. We are Arty Starr, and Margaret-Anne Storey, researchers from the computer-human interaction and software engineering lab, CHISEL, at University of Victoria.*

*This is a follow up interview for the Developer Troubleshooting Study that you were interviewed for last year as part of my PhD research, where we were focused on understanding your experiences of troubleshooting in software development. Since then, we have completed 27 developer interviews and constructed theoretical models of the cognitive experience of Troubleshooting. Our goal today is to see if those models resonate with you, and to discover whether there's any gaps, misunderstandings, or clarifications needed in the models, and to get your feedback and thoughts on how you interpret the models.*

*Only the researchers involved in this study will see your responses, and we will summarize the responses so they are fully anonymized. Your participation in this study is voluntary, and you do not have to answer any question that makes you uncomfortable, and you can stop the interview whenever you want. We'd like to encourage you to refrain from sharing identifying information as much as possible, such as names of 3rd parties, and to refrain from sharing any confidential information during the interview, and we would appreciate it if you respond only based on your personal experiences.*

*This follow up interview should take no longer than 30 min. During the interview we will show you several diagrams that show aspects of the theoretical models, and ask you questions about what you see. I also want to make it clear that it's okay if you disagree or have different ideas, our goal is to improve the clarity of our understanding, so all of this feedback is helpful. If needed, feel free to take your time in answering the questions.*

*So, before we get started, do you have any questions?*

*If this is clear, and you have no more questions, do I have your permission to record the interview?*

**Resonance Testing for Troubleshooting Models**

**1.1. Debugging Context** - So I'm going to be showing you a series of diagrams that together help paint a picture of what is Troubleshooting, and what the cognitive experience of Troubleshooting is like. So first, if you want to just take a look at the diagram, and tell me what you see represented.

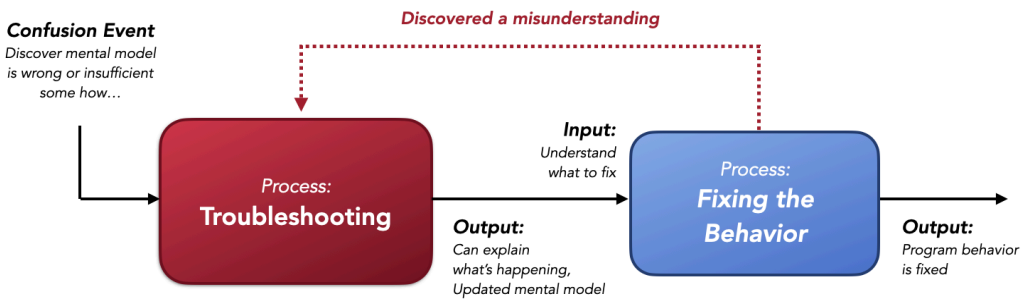

1.2. Do you feel like this resonates with you? Is there anything here that stands out to you that feels inaccurate or that you would change? (If needed, clarify the meaning of resonance, by rephrasing: "Does your experience match what I'm showing you? If it doesn't, I'd like to know...")

1.3. On a scale from 1 to 5, one being it doesn't resonate and feels very different than how you would think about it, and 5 being it resonates and fits with your own mental model, how would you rate this on a scale from 1 to 5?

**2.1. Creation Process Context** - So next diagram, if you want to just take a look and tell me what you see represented...

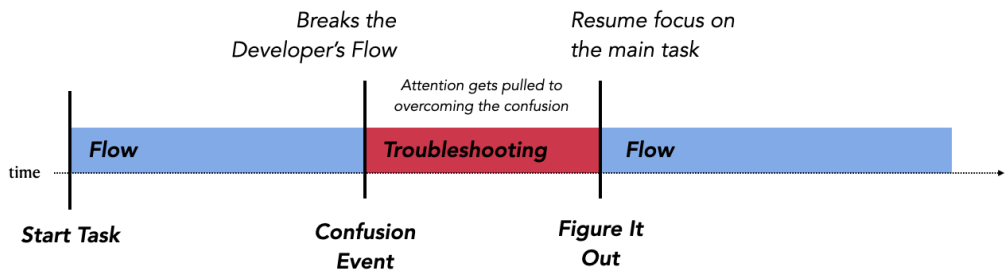

2.2. Do you feel like this resonates with you? Is there anything here that stands out to you that feels inaccurate or that you would change?

2.3. On a scale from 1 to 5, one being it doesn't resonate and feels very different than how you would think about it, and 5 being it resonates and fits with your own mental model, how would you rate this on a scale from 1 to 5?

**3.1. Experiential Feelings** - So next diagram, if you want to just take a look and tell me what you see represented...

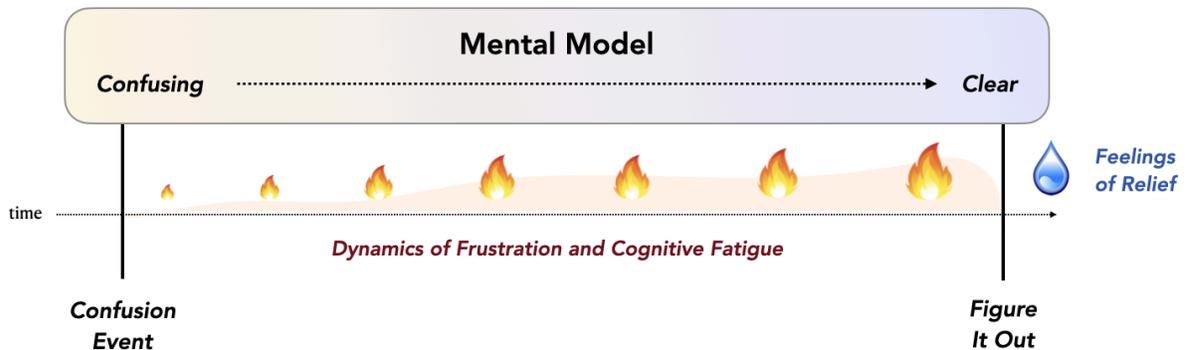

3.2. Do you feel like this resonates with you?  Is there anything here that stands out to you that feels inaccurate or that you would change?

3.3. On a scale from 1 to 5, one being it doesn't resonate and feels very different than how you would think about it, and 5 being it resonates and fits with your own mental model, how would you rate this on a scale from 1 to 5?

**4.1. Intuition Feelings -** So next diagram, if you want to just take a look and tell me what you see represented...

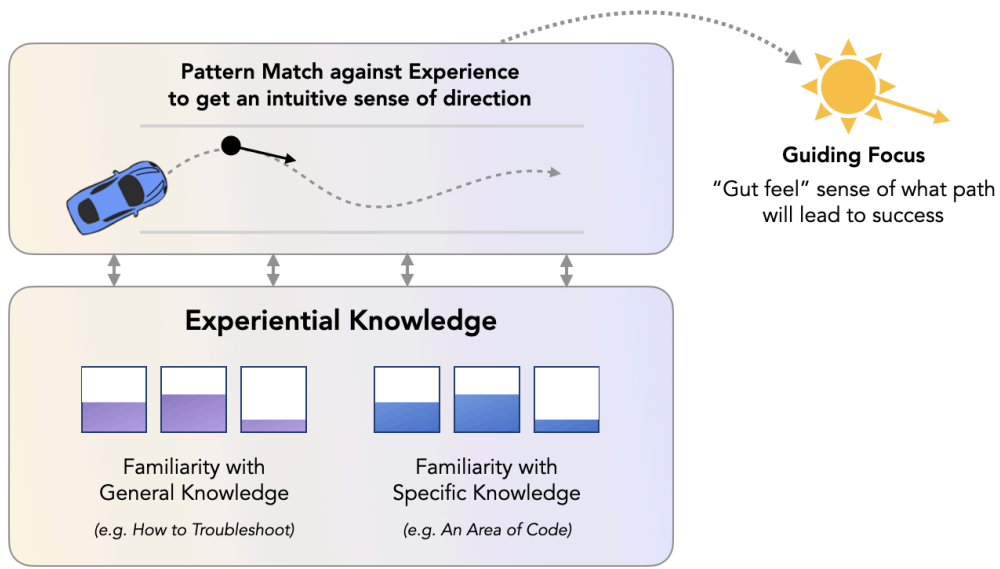

4.2. Do you feel like this resonates with you?  Is there anything here that stands out to you that feels inaccurate or that you would change?

4.3. On a scale from 1 to 5, one being it doesn't resonate and feels very different than how you would think about it, and 5 being it resonates and fits with your own mental model, how would you rate this on a scale from 1 to 5?

**5.1. Striving Process** - So next diagram, if you want to just take a look and tell me what you see represented...

# Troubleshooting: A Process of Trying to Get Clearer

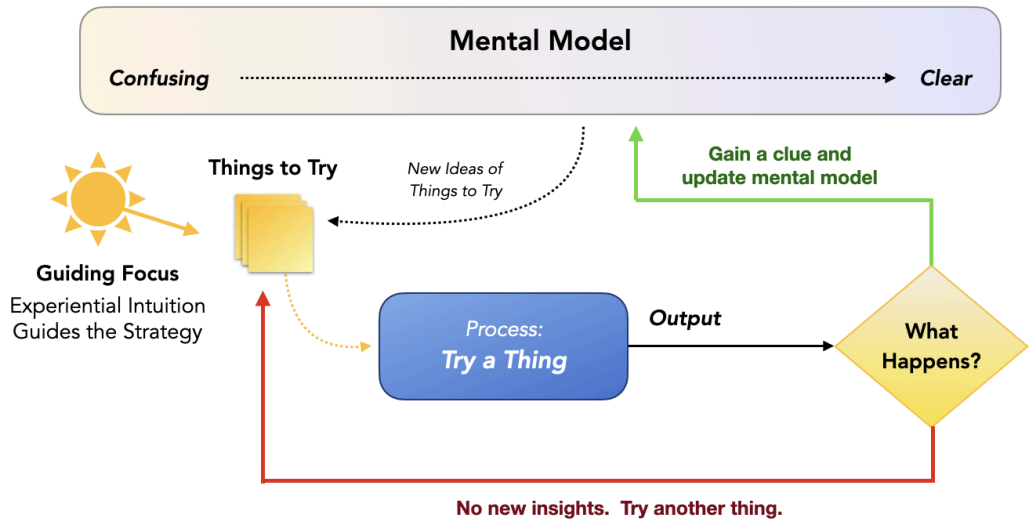

5.2. Do you feel like this resonates with you?  Is there anything here that stands out to you that feels inaccurate or that you would change?

5.3. On a scale from 1 to 5, one being it doesn't resonate and feels very different than how you would think about it, and 5 being it resonates and fits with your own mental model, how would you rate this on a scale from 1 to 5?

**6.1. Actions within Striving Process -** So next diagram, if you want to just take a look and tell me what you see represented...

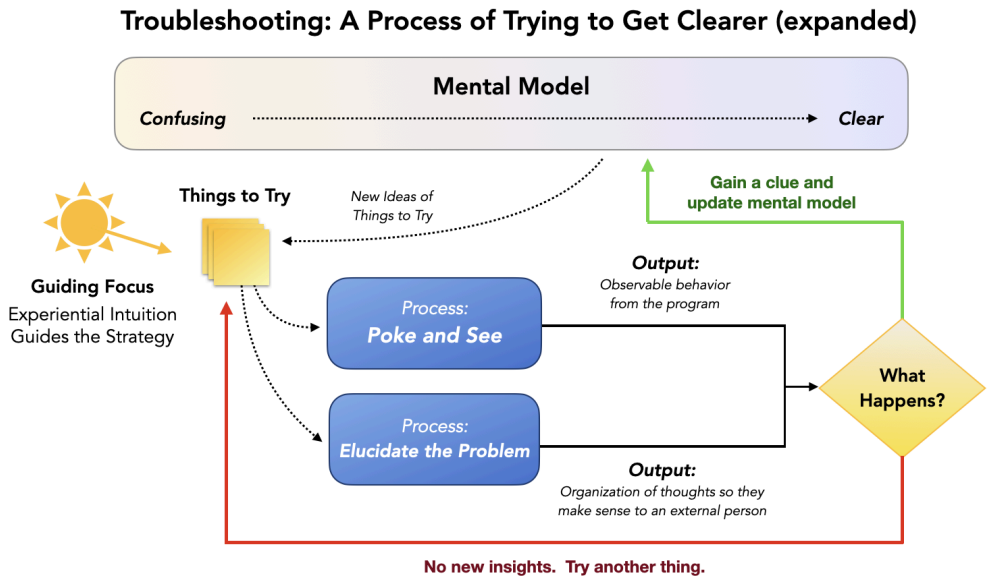

6.2. Do you feel like this resonates with you? Is there anything here that stands out to you that feels inaccurate or that you would change?

6.3. On a scale from 1 to 5, one being it doesn't resonate and feels very different than how you would think about it, and 5 being it resonates and fits with your own mental model, how would you rate this on a scale from 1 to 5?

**Final Reflections**

7. Do you have any final thoughts after seeing all the diagrams, or any other reflections since we met last time?

**Final Wrap Up**

*So that concludes our follow up interview, thank you for participating in the study. We will securely process the recording and the transcript. If at any point in time, you decide you do not want to participate in the study, or if you have any questions, you can email us.*

*Our next step is to refine the models as needed until they resonate with the developers, and to finish the academic paper writeup of the findings. Would you be interested in reading the academic paper for the study when it's available? Would you be open to me emailing you about participating in another short interview if it's necessary to complete the study? For example, if we do another iteration on the diagrams you saw today, and we needed to get more feedback.*

*Do you have any final questions for us or any further comments?*

*Alright, thank you very much for your time, and have a great day.*



\end{document}